\documentstyle[preprint,aps]{revtex}

\begin{document}
\draft

\renewcommand{\thesection}{\arabic{section}}

\title{A Comprehensive Dynamical Study of Nucleation \\ and Growth in a 
One--Dimensional Shear Martensitic Transition}

\author{B.P.~van Zyl and R.J.~Gooding}
\address{Department of Physics, Queen's University,\\ Kingston, Ontario K7L 3N6,
Canada}
\date{\today}

\maketitle 
\begin{abstract}

We have constructed a complete hydrodynamic theory of nucleation
and growth in a one--dimensional version of an elastic shear martensitic 
transformation with open boundary conditions where we have accounted
for interfacial energies with strain--gradient contributions. We have studied 
the critical martensitic nuclei for this problem: Interestingly, 
the bulk critical nuclei are {\em twinned} structures, 
although we have determined that the dominant route for the formation of 
martensite is through {\em surface nucleation}. We have analytically solved 
for the surface nuclei and evaluated exact nucleation rates showing 
the strong preference for surface nucleation.
We have also examined the growth of martensite: 
There are two possible martensitic growth fronts, {\em viz}.,
dynamical twinning and so-called two--kink solutions.  These
transformation fronts are separated by a {\em dynamical} phase transition.  
We analytically derive this phase diagram and determine expressions for 
the speeds of the martensitic growth fronts.

\end{abstract}
\newpage

\section{Introduction and Motivation:}
\label{sec:motivation}

Nature offers a variety of transformations between different crystallographic
states.  For many crystals, the result of such a phase transition involves
a dramatic change in the macroscopic shape of the material.  
Quite often, associated with these macroscopic changes is the formation of
characteristic domain patterns, sometimes called microstructure.
When viewed under a transmission electron microscope (TEM), the 
most commonly observed microstructures are those related to twinning, 
{\em viz.}, symmetry related variants of the product phase(s) oriented in a 
characteristic pattern. 
The study of these materials is very desirable because many technologically 
important compounds, such as shape memory alloys and the A15 superconductors, 
undergo what are commonly known as martensitic transformations ---  
these transformations lead to the formation of the microstructre 
discussed above \cite {salje}.  

Martensites are to be distinguished from other materials undergoing structural
phase transitions by the non--diffusive nature of the transition.  To be 
specific, 
(i) if the structural transformation occurs without a net diffusion of 
atoms across the crystalline unit cell boundary, (ii) the transition
is discontinuous (also referred to as first--order), and (iii) the transformation
involves shear strains, the transformation is said to
be martensitic, and the resulting material a martensite \cite {barsch_krum1}.  
These transformations can always be described in terms of strains and 
possibly lattice modulations (or so-called lattice vibrational modes) 
\cite {lee_oakland}.  Here we shall specialize to a subclass of martensitic 
transitions known as proper ferroelastic transitions for which the only
relevant variables are the purely elastic strains.

Martensitic transformations occur by a nucleation and growth processes, 
and the {\em dynamical} path by which such transformations proceed is the 
subject of this paper. We explore the
dynamical evolution of a phenomenological model of a one--dimensional,
purely elastic, martensitic transformation based on the concept of a strain
order parameter.  We determine both the nucleation and dynamical growth 
aspects of the transformation for this simple model. We also make contact
with the formation of microstructure, something that Bales and one of us 
\cite {b/g} has recently proposed can be associated with the dynamics of
martensitic growth.

The consideration of a lower dimensional model has many advantages:  
Firstly, many of the practical problems ({\em e.g}., numerical integration of
the equation of motion) associated with higher dimensional systems are 
eliminated. Secondly, lower dimensional systems often allow for the possibility
of obtaining exact analytical results. We shall indeed find many exact 
analytical solutions, and hopefully these results will be able
to be extended into higher dimensions in the future.

Summarizing our main findings: \hfill\break
{\bf I} -- One may numerically solve for the
bulk critical nucleus for such transformations, and then obtain
an excellent analytical approximation to this structure. {\em The bulk
critical nucleus is twinned}. \hfill\break
{\bf II} -- For any finite system with open boundary conditions,
the critical nucleus may be solved for exactly, and we find that
it exists {\em at the surface} of the system. It's energy is (almost
exactly) four times less than that of the bulk nucleus (namely, the
critical nucleus for a system of infinite extent), and it is not
twinned. \hfill\break 
{\bf III} -- As shown previously \cite {b/g}, interesting domain wall motions
are associated with the growth of martensite when the system is quenched
to temperature below the transition temperature.
Here we will show that one can understand the variety of growth fronts
that are found for these dynamics by analytically deriving
interfacial growth speeds for a variety of domain walls, and then applying
a {\em local} stability analysis. \hfill\break

We believe that apart from
the inclusion of thermal fluctuations, something that would require
simulations based on, {\em e.g}., Langevin dynamics, this work represents as 
complete a numerical plus analytical examination of this one--dimensional
problem as is possible.

Our paper is organized as follows.  In Section 2 we provide all of the
requisite mathematical formalism to understand our one--dimensional model 
system.  Then, in
Section 3 we determine the critical nucleus for both bulk and surface
nucleation, with particular attention paid to the intimate relation between 
the bulk and surface solutions.  Section 4 provides a
comprehensive study of the growth dynamics that can occur for
initial states that are supercritical.  We also provide analytical work
that successfully explains much of the
observed numerical results, and allows for the exact derivation
of various dynamically interesting quantities ({\em i.e.}, the growth speed
of a transformation front).  We have tried to compare, wherever possible, 
our numerical results to those of relevant experiments. Lastly, in
Section 5 we summarize our results, and forecast the success of
similar phenomenologies to the more interesting case of higher dimensional
systems.

\medskip
\medskip
\section{Formalism and the Equation of Motion:}
\label{sec:formalism}

In this section we will present the formalism necessary to describe a 
model of a one--dimensional (1D) version of a first--order, elastic shear 
transformation via a Ginzburg--Landau (GL) theory. This model 
was first applied to martensite by Falk \cite {falk}, and we believe
that it is the simplest GL potential depending on only one spatial variable
that includes any of the important characteristics of these transitions.
For example, we wish to focus
on martensitic transitions wherein the high--temperature parent phase
has a sufficiently high symmetry that the transformation shear strains lead
to degenerate martensitic product states, something that is a prerequisite
for the formation of twinning --- we use Falk's GL potential since it 
possesses this feature.

Falk's mean--field theory of martensitic phase transitions lies in the
construction of the phenomenological GL free--energy density.
To be specific, one considers a displacement field $u(x,t)$ and the associated
strain $e(x,t) = \partial u / \partial x \equiv \partial_{x} u$. Then, one imagines that the local
free energy density, $f_L$, of the system can be characterized by a nonlinear function of the form:
\begin{equation}
f_L(e) = \frac{1}{2} A~\delta T e^{2} - \frac{1}{4} B e^{4} +\frac{1}{6} C e^{6}~~~. 
\label{eq:localf}
\end{equation}
In Eq.~(\ref{eq:localf}), $A, B,$ and $C$ are 
positive, temperature independent, phenomenological constants
and $\delta T = T - T_{c}$. The quantity $T_{c}$ represents the temperature
at which the unstrained parent phase ($e=0$) becomes unstable.
In a linear theory, the undercooling, $\delta T$, would be related
to the linear elastic constant $c$ of the high--temperature parent
phase by $c = A~\delta T$.

The first three terms in the Landau energy density above represent the
local response of the system to a given strain.  
The particular symmetry of the local potential, {\em viz}. $f_L(e) = f_L(-e)$, 
ensures that both positive
and negative shear strains of equal magnitude have the same energy.
For 
an appropriate choice of model parameters, a local free energy density
similar to the one illustrated
in Fig. \ref{fig:potentialfig} can be produced. GL 
densities of this form can be attributed
to systems for which the (high--temperature) parent phase 
and (low--temperature)
product phase(s) are separated by a first--order, elastic shear transformation.
In our model, temperatures greater than the first--order       
transition temperature, $T_{1}$, lead to only 
one absolute minimum, $e = 0$, on
the energy surface; this strain state corresponds to the
parent phase.
Below $T_{1}$, the parent state becomes metastable and two doubly 
degenerate, stable minima develop at $e = \pm e_{m}$ 
(the so--called martensitic strains).
These minima correspond to symmetry related variants of the product phase
and are often refered to as martensitic twins.

We also wish to include so--called non--local elastic forces. 
For the one--dimensional problem that we are considering,
the appropriate form of this energy is
\begin{equation}
f_{NL} = \frac{1}{2} D (\partial_{x} e)^{2}~~~.
\label{eq:nonlocalf}
\end{equation}
Terms are of this form are responsible for many properties. Firstly, they 
account for
non--local forces associated with inhomogeneous strain fields. Secondly,
in the context of phase--transition theory, such terms represent the 
domain--wall energy associated with the inhomogeneities of two--phase regions
(this is the analogue of the so--called Ginzburg energy found in
the theory of type--II superconductors).
Lastly, these terms break the scale invariance implicit in Eq.~(\ref{eq:localf})
for a bulk system --- we shall elaborate on this last feature below.

The total elastic free energy density is thus $f_L + f_{NL}$, implying that for
a system of length $L$ defined by the spatial range 
${- L \over 2} \leq x \leq {L \over 2}$ the total elastic free energy is
\begin{equation}
F = \int_{{- L \over 2}}^{{L \over 2}} [~f_L (x) + f_{NL} (x) ~] dx~~~.
\label{eq:potential}
\end{equation}

In order for the equation of motion to be specified for this system,
one could make use of the time--dependent Ginzburg--Landau
theory (TDGLT) to construct a first--order in time, nonlinear, nonlocal
partial differential equation (PDE) for the dynamical
evolution of the system \cite{chan}. This
approach, however, has been shown to lead to 
a completely incorrect description
of the {\it dynamical} growth aspects of elastic--shear 
transformations \cite {b/g}.  Instead, one must properly account for the
hydrodynamic character of the sound waves through the inclusion of the
kinetic energy density associated with the propagating growth interfaces.
This approach makes use of the fact that nonlinear elastic models of
the type discussed here can be treated as nonconvective, hydrodynamic
systems \cite {r/g_openbcs}. 

Propagating disturbances in a solid, {\em e.g.}, thermal phonons,   
necessarily involve the displacement of a finite mass of material and hence
contribute to the total mechanical energy of the system.  Recalling that the
displacement field, $u(x,t)$, describes the physical displacements of the atoms
in a solid relative to some chosen undistored system, the kinetic
energy, $T$, is given by
\begin{equation}
T = \frac {1}{2} \rho \int_{{- L \over 2}}^{{L \over 2}} [~(\partial_{t} u)^{2}~] dx~~~,
\label{eq:kinetic energy}
\end{equation}
where $\rho$ is the linear mass density of the undistorted bar.
In the spirit of our hydrodynamic considerations, we also include a
Rayleigh dissipation function, $R$, with sound wave viscosity $\gamma$, {\em viz}.,
\begin{equation}
R = \frac {1}{2} \gamma \int_{{- L \over 2}}^{{L \over 2}} [~(\partial_{t} \partial_{x} u)^{2}~] dx~~~,
\label{eq:dissipation}
\end{equation}
that allows for the dissipation of energy of the sound waves \cite {Landau}.
It is to be stressed that accounting for the damping in this manner
ensures that the sound waves are always propagating at sufficiently long 
wavelengths; this is the sense in which the system's hydrodynamic character 
is being properly accounted for \cite {martin}.

With these ingredients, the equation of motion plus the boundary conditions
may be determined precisely as described, {\em e.g.}, in Ref. \cite {r/g_openbcs} --- we stress
that we are studying a finite system, and thus the boundary conditions
are an important part of this problem.  The bulk equation of motion for the system is
\begin{equation}
\rho (\partial^{2}_{t} u) = (\partial^{2}_{x} u) [A \delta T - 3 B (\partial_{x} u)^{2}
+ 5 C (\partial_{x} u)^4] - D (\partial^{4}_{x} u) + \gamma (\partial_{t} \partial ^{2}_{x} u)~~~,
\label{eq:bulk equation}
\end{equation}
and the four boundary conditions are
\begin{equation}
A \delta T (\partial_{x} u) - B (\partial_{x} u)^{3} + C (\partial_{x} u)^{5} +
\gamma (\partial_{x} \partial_{t} u) - D (\partial^{3}_{x} u) = 0
~~$at$~~~~~ x = \pm {L\over 2}
\label{eq:bc1}
\end{equation}
and
\begin{equation}
(\partial^{2}_{x} u) = 0~~~$at$~~~ x = \pm {L\over 2}~~~.
\label{eq:bc2}
\end{equation}
By differentiating Eq.~(\ref{eq:bulk equation}) once, 
these equations can also be conveniently restated in terms of the shear strain:
\begin{equation}
\rho (\partial^{2}_{t} e) = \partial^{2}_{x} [A \delta T e - B e^{3} +
C e^{5} - D (\partial^{2}_{x} e) + \gamma (\partial_{t} e)]~~~,
\label{eq:eom strain}
\end{equation}
with boundary conditions
\begin{equation}
(\partial_{t} e) = \frac {1}{\gamma} (D (\partial^{2}_{x} e) - A \delta T e + B e^{3} - C e^{5})
\label{eq:strainbc1}
\end{equation}
and
\begin{equation}
(\partial_{x} e) = 0,~~~$at$~~~x = \pm {L\over 2}~~~
\label{eq:strainbc2}
\end{equation}
In certain instances (see below) it will be advantageous to refer to the 
constitutive equations in this form.

These equations involve a large number (eight) of material parameters.
However, we can substantially reduce the number of parameters required to
model our system through scaling analysis, and the result of this analysis
for bulk systems has been discussed elsewhere \cite {b/g,r/g_saddle}.
Here we are considering systems of a finite length, and have chosen to
rescale our dynamical equations through the use of a generalized dimensional 
analysis for boundary valued problems \cite {Bluman}. The mathematical details
of this are summarized in the Appendix A; the physics behind the resulting
equation of motion and boundary conditions are as follows: There
are four dynamical units, {\em viz}., distance, time, mass and temperature. Rescaling
the ``length" of each of these four dynamical units allows us to eliminate
four of the material parameters. Further, we can scale the two parameters
depending only on lengths, {\em viz}. $x$ and the displacement field $u$, using different
scale factors (this is analogous to simply rescaling the unit strain), and
thus we find that we can reduce our problem down to one involving only three
material parameters, $\Lambda$, the scaled mass density, $\tilde{\delta T}$,
the scaled undercooling, and ${\tilde L}$, the scaled length of the system.
The resulting equation of motion and boundary conditions are
\begin{equation} 
\Lambda (\partial^{2}_{t} u) = (\partial^{2}_{x} u) 
[\delta T - 3 (\partial_{x} u)^{2} + 5 (\partial_{x} u)^4] 
- (\partial^{4}_{x} u) + (\partial_{t} \partial ^{2}_{x} u)~~~,
\label{eq:rescaled eomu}
\end{equation}
\begin{equation}
\delta T (\partial_{x} u) - (\partial_{x} u)^{3} + (\partial_{x} u)^{5} +
(\partial_{x} \partial_{t} u) - (\partial^{3}_{x} u) = 0,
~~$at$~~~x = \pm {L\over 2}~~~,
\label{eq:rescaled bc1}
\end{equation}
\begin{equation}
(\partial^{2}_{x} u) = 0, ~~~$at$~ x = \pm {L\over 2}~~~,
\label{eq:rescaled bc2}
\end{equation}
(where in the above system of equations, and from now on, we drop the tildes).

Further, in terms of the scaled variables the local potential is now given by
\begin{equation}
f_L(x) = \frac{1}{2} \delta T e^{2} - \frac{1}{4} e^{4} +\frac{1}{6} e^{6} 
\label{eq:localf_scaled}
\end{equation}
and thus the martensitic strains are now given by
\begin{equation}
e_m (\delta T) = \pm \sqrt{\frac{(1 + \sqrt{1 - 4 \delta T})}{2}}~~~.
\label{eq:emart}
\end{equation}
Thus, one has the following sequence for the relative stability of
the unstrained and martensitic states \cite  {falk}: 
$\delta T \geq 1/4$, only the unstrained state is locally stable; 
$3/16 \leq \delta T \leq 1/4$, the unstrained state is locally stable
and the doubly degenerate martensitic states are metastable; 
$0 \leq \delta T \leq 3/16$, the unstrained state is metastable and
doubly degenerate states are stable;
$\delta T \leq 0$, only the doubly degenerate martensitic states are locally 
stable. 

\medskip
\medskip
\section{The Critical Nucleus:}
\label{sec:nucleation}

In the model discussed above, we have tacitly assumed that the system
is coupled to an infinite heat bath so that the dynamics can be considered
to be {\em isothermal}.  However, if the system is found in the metastable 
unstrained state at temperatures such that $\delta T < 3/16$,
a local fluctuation (of the displacement field) can
lead to the subsequent formation of a locally stable region of martensite
which can grow to expel (sometimes) all of the unstrained state. 
That is, the decay of the metastable state ({\em e.g.}, the unstrained bar)
requires an excitation called the ``critical nucleus" with finite activation
energy $\Delta E$.  Any initial displacement
profile that lies ``below" this saddle--point configuration will decay to zero, whereas
states lying ``above" this profile may be able to escape from the basin of
attraction and grow to the (globally) stable product phase.
In this
section we shall describe both the novel twinned nucleus, which characterizes
the bulk nucleus of a spatially infinite system, and the localized (in strain)
surface states, which correspond to the true critical nuclei (saddle--points)
of this problem.  Then, in the next section
we will provide some examples of the fascinating growth phenomena that can
result as the system approaches steady state.

We can find the critical nucleus by solving the following nonlinear ODE
plus nonlinear boundary conditions:
\begin{equation}
\partial^{4}_{x} u - \partial^{2}_{x} u (\delta T - 3 (\partial_{x} u)^{2} +
5 (\partial_{x} u)^{4}) = 0~~~,
\label{eq:bulk zeroforce}
\end{equation}
\begin{equation}
\delta T (\partial_{x} u) - (\partial_{x} u)^{3} + (\partial_{x} u)^{5} -
(\partial^{3}_{x} u) = 0~~~~$at$~x =  \pm \frac{L}{2}~~~,
\label{eq:bc1 zeroforce}
\end{equation}
\begin{equation}
(\partial^{2}_{x} u) = 0~~~~$at$~x =  \pm \frac{L}{2}~~~.
\label{eq:bc2 zeroforce}
\end{equation}
The above system of nonlinear equations are derived from considering the
static, zero--force case of 
Eqs.~(\ref{eq:rescaled eomu},\ref{eq:rescaled bc1},\ref{eq:rescaled bc2}).   
The bulk or ``saddle--point'' critical nucleus that we are interested in, 
to be denoted by $u_{bcn} (x)$, is the lowest energy, {\em localized}
configuration satisfying the above set of equations --- the localization of the 
displacement field is a consequence of the physical constraint that for a 
bulk nucleus, the boundaries must be unaffected by the spatial 
perturbation. (We stress that this constraint is based on physical 
considerations, and does not arise from a purely mathematical treatment 
of the formalism presented in the previous section.)  

\subsection{Bulk Solutions:}
\label{subsec:bulk nucles}

In this section we consider a system of infinite extent, thus
precluding the possibility that the system nucleates at a boundary.
For such a bulk critical nucleus, 
Eqs.~(\ref{eq:bc1 zeroforce}) and (\ref{eq:bc2 zeroforce}) are irrelevant,
and the solutions of interest must solve Eq.~(\ref{eq:rescaled eomu})
and satisfy $\partial^{j}_x u = 0$ as $x \rightarrow \pm \infty$ 
for $j = 0, ..., 4$.

The work required to find the critical nucleus is greatly simplified
by the observation that the equations determining it depend on only 
one (scaled) material parameter, {\em viz}., $\delta T$. At this point we
choose to refer to the ratio of energies $E_{W}/E_{B}$ instead of $\delta T$, 
where $E_{B}$ is the energy barrier separating the unstrained state from the 
martensitic wells, and $E_{W}$ is the energy difference between the
$e = 0$ and $e = \pm e_m$ wells --- these two energies are displayed in
Fig. 1. In terms of $\delta T$
\begin{equation}
{E_{W} \over E_{B}} (\delta T) = 
{1 - 6 \delta T + (1 - 4 \delta T) \sqrt {1 - 4 \delta T} \over
- 1 + 6 \delta T + (1 - 4 \delta T) \sqrt {1 - 4 \delta T}}
\label{eq:ewvseb}
\end{equation}
and from now on we shall refer to this ratio, quite simply because
it better distinguishes between the differing
``shapes" of the local Landau potential Eq.~(\ref{eq:localf}).
Noting that $\delta T = 1/6~\Rightarrow~E_{W} = E_{B}$, we have
chosen to examine $E_{W}/E_{B}$ = 0.2, 1.0, and 5.0, corresponding
to $\delta T$ = 0.1823, 1/6, 0.1295.

We an easily determine one feature of the critical bulk nucleus for
this system --- symmetry arguments plus the physical constraint of 
a vanishing displacement field at the boundaries allow us to specify 
that the bulk critical nucleus is a localized, symmetric displacement field.
Namely, since localized symmetric states are lower in energy
because there are fewer domain walls present, we know that this must be the
symmetry of our bulk critical state. 

We have solved for the bulk critical nuclei $vs.~E_{W}/E_{B}$ numerically.  
This is a difficult problem since apart from the symmetry of
the solution, one does not know {\em a priori} where in function space to 
begin one's search.  In order to
minimize this difficulty, we have followed an approach based on the 
full dynamical equations.  
We assume that the critical nucleus should be a configuration approximately
of the form 
\begin{equation}
u(x) = u_{0}~${\rm exp}$(-x^{2n}/d^{2n})~~~. 
\label{eq:u_interpolate}
\end{equation}
We consider this functional form because it smoothly interpolates between 
a classical nucleus $(n \rightarrow \infty)$ and a Gaussian fluctuation
of the displacement field with broad domain walls $(n=1)$ --- these two limiting
states are shown in Fig. \ref{fig:limiting_nuclei}.  We then specify that
$d$ is not too narrow, say $d = 10$ domain wall lengths (see, {\em e.g.}, Appendix B),
and for simplicity take $n=1$.  Thus, for the static displacement field
\begin{equation}
u(x) = u_{0}~${\rm exp}$(-x^{2}/d^{2})
\label{eq:ansatz}
\end{equation}
at $t=0$, we numerically integrate
forward in time under the dynamics of the system to find the ``critical"
value of $u_{0}$ as a function of $E_{W}/E_{B}$, {\em viz}., the value of $u_{0}$
for which perturbations above or below will grow or decay respectively.
In this way, we have {\em dynamically} eliminated those initial configurations
which could not possibly be candidates for the critical nucleus.

Having obtained this reduced function space, 
a numerical solution for the critical nucleus becomes tractable: 
we employ a numerical relaxation method
\cite{numerical_recipes} that solves the of zero--force equation
(plus boundary conditions) using as input Eq.~(\ref{eq:ansatz})
and the critical value of $u_0$ found as described above. 
These bulk critical nuclei for the above--mentioned 
ratios of $E_{W}/E_{B}$, are shown in Fig.~\ref{fig:critical_nucleus}.  
(We stress that these solutions are found for $L$ being
sufficiently large such that if the length of system is then doubled, our 
solutions are unchanged --- this confirms that these are indeed
bulk nuclei unaffected by the surfaces.)
It is interesting to observe that {\em the critical nucleus is a fully
twinned strain state}. This may be understood from Fig.~3b where
the strain profile for $E_{W}/E_{B} = 1.0$ is displayed.
This behaviour is a direct consequence of the physical as opposed to mathematical 
constraint of a vanishing displacement field at the boundaries, {\em i.e.}, 
a localized, twinned strain state allows for the localization of 
the displacement field. (For example, in any configuration with only one sign
of strain throughout the entire system, at least one boundary must
be displaced.)

To further check that these numerical solutions behave like saddle--point
solutions, we have examined their evolution under the full dynamical equations
of motion in two ways: 
\begin{equation}
{\rm (i)}~~~~~u_{bcn}(x) \rightarrow (1 \pm \epsilon)~u_{bcn}(x)
\label{eq:nucl test 1}
\end{equation}
\begin{equation}
{\rm (ii)}~~~~~u_{bcn}(x) \rightarrow u_{bcn}((1 \pm \epsilon)x)
\label{eq:nucl test 2}
\end{equation}
\noindent
where numerically we have used $\epsilon$ values as small as 0.001.
In the case of (i), it was found that the $1 - \epsilon$
configuration was always subcritical, while the $1 + \epsilon$  configuration
was always supercritical.  Further, it was also found that a scaling of 
the independent variable described in (ii) was supercritical (subcritical)
if the state was increased (decreased) in size. Thus, these
critical nuclei have just enough driving force (energetically
speaking) and are just big enough to drive the system into the martensitic
state.

As the ratio $E_{W}/E_{B} \rightarrow \infty$ ({\em viz}., as $\delta T
\rightarrow 0$), it becomes increasingly difficult to
obtain converged saddle--point solutions to the zero--force equations.  
We find that the critical nucleus becomes more non--classical,
{\em viz.}, more bell--shaped; eventually, it becomes too numerically sensitive to
relax to a saddle--point solution.  However, in the opposite limit
of $E_{W}/E_{B} \rightarrow 0$,
the shape of the critical nucleus approaches that of its
classical analogue, {\em viz}., the step function found from $n \rightarrow \infty$
in Eq.~(\ref{eq:u_interpolate}), and for this limit we have found that it 
is possible to produce an excellent approximation to the bulk critical 
nucleus from purely analytical considerations. The details 
of this analysis are provided in Appendix B, and here we summarize our 
results.

We consider the zero--force equations now in terms of the strain variable $e$.
Firstly, we focus on the equations for an infinite system with the constraint 
that 
$\partial^{j}_{x} e = 0$ at $x = \pm \infty$ and $j =0, 1, .., 4$.  
One then finds the solutions
\begin{equation}
e_{0}(x) = \pm 2 \sqrt{6 \delta T} \sqrt{\frac{{\rm exp}
[2 \sqrt{\delta T} (x - x_{0})]}{3 - 16 \delta T 
+ 6~{\rm exp}[2 \sqrt{\delta T} (x - x_{0})] +
3~{\rm exp}[4 \sqrt{\delta T} (x - x_{0})]}}
\label{eq:saddle_strain}
\end{equation}
where $x_{0}$ is some integration constant.  Each of these solutions correspond
to a strain field that is symmetric about $x = x_{max}$ where
\begin{equation}
x_{max} = x_0 + \frac{1}{4 \sqrt{\delta T}} {\rm ln}
\Big(\frac{3 - 16 \delta T}{3}\Big)
\label{eq:xmax}
\end{equation}
Now note that since $e(x)$ is a symmetric function, we know that
the boundaries of the system must be displaced --- put another way,
the strain field defined in Eq.~(\ref{eq:saddle_strain}) is not
a localized solution.
Figure \ref{fig:soliton} shows a plot of 
Eq.~(\ref{eq:saddle_strain}) for $E_{W}/E_{B} = 1.0$. 
A graphical comparison with the exact localized solutions, {\em viz}. those 
found from the relaxation technique [see Fig. 3b], suggest that each 
of the $\pm e_{0}(x)$ strain configurations correspond to one half of 
the strain strain states that make up the twinned bulk critical nucleus.  
We have tested this conjecture and found that
a {\em linear} combination of the form
\begin{equation}
e(x) = e_{0}(x-x_{0}) - e_{0}(x+x_{0})
\label{eq:linear combo}
\end{equation}
(which by the even symmetry of Eq.~(\ref{eq:saddle_strain}) is an odd
function, the same symmetry as our numerically determined critical
nuclei) does in fact satisfy the bulk zero--force equation 
{\it with~a~small~residual~error}. This error is a function of $x_0$,
a quantity which remains to be determined.

Then, we use this approximation in the following way: if we desire
an analytical approximation to our (exact) numerical result, we
can find an $x_0$ such that Eq.~(\ref{eq:linear combo}) is as close
as possible to the exact strain profile. Thus, define the residual
error to be the integral of the square of the difference over
the length of the system, and then minimize this quantity with respect
to $x_0$.  For small $E_{W}/E_{B}$,
this procedure leads to superb agreement with our (numerically) exact solutions.
Figure \ref{fig:approx. strain nuclei} illustrates the somewhat amazing success
of this procedure for $E_{W}/E_{B} = 1.0$.
At this and lower temperatures we thus find that this function
yields an excellent fit to the exact solution shown in Fig. 3b.
Since we cannot find analytical solutions for $E_{W}/E_{B} \gg 5.0$, this
procedure is limited to low values of this ratio.

To understand the success of this approximation, we note that
although the nonlinear nature of this problem excludes the possibility of 
{\em any} linear combination of zero--force solutions exactly satisfying 
Eq.~(\ref{eq:bulk zeroforce}), such as we have used in Eq.~(\ref{eq:linear combo}),
the structures that we are superimposing on one another are like two solitons.
The solitons are particle--like entities, and it is usual in soliton
theory to ascribe an interaction energy between such particles. 
Usually \cite {rajaraman},
this interaction decays as an exponential. Thus, in Eq.~(\ref{eq:linear combo})
we have a soliton interaction energy that goes as $\exp (-2 x_0)$. As shown in
Fig. 3b, for the ratios of $E_{W}/E_{B}$ studied here, the value of $x_0$ is
quite large, and thus this interaction is very small. This simple consideration
explains the success of our analytical work in producing an approximate 
bulk critical nucleus.

\subsection{A Critical Nucleus at the System's Surface:}
\label{subsec:surface nucleus}

In the previous subsection the bulk critical nuclei were found.
We wish to stress that in deriving these configurations
the boundary conditions given in Eqs.~(\ref{eq:bc1 zeroforce},\ref{eq:bc2 zeroforce})
were effectively ignored due to our consideration of a system
of infinite extent. Now we include the boundaries by focusing on
a system that is finite in at least one direction. We will show that
the bulk critical nuclei are {\em not} critical nuclei for these
systems --- in fact, we shall show and analytically derive that
the critical nuclei for systems that include at least one free surface
always possess critical nuclei {\em localized at the surface}. Such
states are found to have a significantly lower energy than the bulk
critical nuclei.

We discovered these solutions quite easily --- simply note that localized 
surface states in strain space follow directly from the analytical work on 
the bulk critical nucleus discussed in Appendix B and explicitly
stated in Eq.~(\ref{eq:saddle_strain}).  
Indeed, if the localized, single--humped bulk solution
$e_0(x - x_{0})$ is centered at one of boundaries, $x_{max} = \pm L/2$, 
the resulting state satisfies the boundary conditions
and provides a stationary solution to the bulk equation, namely 
Eq.~(\ref{eq:bulk equation}). Since such a state is similar
to that in shown in Fig. 3b and only involves non--zero strains in
one fourth of the space that a bulk nucleus exists over, clearly it possesses
an energy that is (almost exactly) only one quarter of the energy of 
the critical nuclei solutions that we found for the bulk.

To see that this is indeed a saddle--point solution for a finite
system with free boundaries, we employ a hydrodynamic approach \cite {r/g_saddle}.
As in any hydrodynamic theory with a conserved quantity (in this case, momentum), 
a natural physical interpretation of the equation of motion is that of a continuity 
equation, {\em viz}.,
\begin{equation}
\partial_{t}(\Lambda \partial_{t} u) = \partial_{x} J~~~,
\label{eq:continuity eq}
\end{equation}
where $J$, the one dimensional momentum current density of the system, is given by
\begin{equation}
J = \delta T (\partial_{x} u) - (\partial_{x} u)^{2} + (\partial_{x} u)^{3} +
(\partial_{t} \partial_{x} u) - (\partial_{x}^{3} u)~~~.
\label{eq:momentum current}
\end{equation}
For the case of static, zero--force solutions, the continuity equation
dictates that the (one--dimensional) divergence of the momentum current, 
$\partial_x J$, must be zero, implying that $J$ is a constant. For any
solution that is localized in space, such as Eq.~(\ref{eq:saddle_strain}),
it trivially follows that in fact $J = 0$ ({\em e.g.}, simply consider
a region for which the strain is vanishingly small and thus the constant
$J$ must also vanish). Now note that the nonlinear boundary
condition in Eq.~(\ref{eq:bc1 zeroforce}) is nothing more than
the condition that $J = 0$ at $x = \pm L/2$, and thus this boundary
condition is satisfied. The second boundary condition, 
Eq.~(\ref{eq:bc2 zeroforce}), is satisfied since we have placed the maximum 
of the strain bump at the system's surface.

The natural question that arises is: which of the above--discussed
nuclei is the potent nucleus for a large but finite system?
To answer this question, recall the common definition of the nucleation rate,
$r$, as the ratio of the probability flux $j$ across the saddle point to
the probability $n$ to be in the metastable well: $r = j/n$ ({\em e.g.}, see
the discussion in Ref. \cite{Talkner}). 
We can readily provide the nucleation rate $r$ of a kink at the surface, {\em viz}.,
\begin{equation}
r = S~$\rm{exp}$(-\beta \Delta E/4)~,
\label{eq:nucleation rate}
\end{equation}
where $\Delta E$ is the activation energy for the bulk nucleus. 
(The prefactor $S$ in front of the Arrhenius term is a rate constant.)
The lower activation energy of a nucleus at the
boundary compared to that of a bulk nucleus (as mentioned
above, this ratio of energies is very nearly exactly 1:4) implies that
surface nucleation will always be preferred over homogeneous bulk nucleation,
and thus will certainly be the dominant route to the formation of
martensite. (Of course, the other common inhomogeneity besides a free surface would
be impurities, and these would be expected to compete with the surface
for the role of ``most potent" nucleation centre.)

\medskip
\medskip
\section{The Propagating Dynamics of Martensitic Growth Fronts:}
\label{sec:growth}

In this section we will describe and examine the dynamical growth aspects
of those states which have overcome the nucleation barrier, {\em viz}., states
that lie ``above'' the critical nucleus, as they approach their steady
state profiles.  We will thoroughly explain the observed dependence 
of the dynamical evolution of the product phase, {\em viz}., propagating 
martensitic growth fronts, on the density $\Lambda$, 
the undercooling $\delta T$ and the length of the system, $L$ --- this
complements the brief outline of results for
this phenomenon given in Ref. \cite {b/g}.
In these
investigations we have numerically obtained the evolution of the system under 
consideration by using a variant method of lines \cite{y/g} --- this
provides a high accuracy integration of the equation of motion, {\em viz}., 
Eq.~(\ref{eq:bulk equation}), as well as fully accounting for
the time--dependent, nonlinear boundary conditions given in 
Eqs.~(\ref{eq:bc1},\ref{eq:bc2}).

To be specific, to investigate the dynamical evolution of martensite,
we first specify the parameters characterizing the system, {\em viz.},
the scaled temperature, length and mass density of the system.
(To allow us to focus on a physically relevant range of the scaled 
mass density parameter, we note that
recent experimental work on the purely elastic bcc $\rightarrow$ fcc
transformation of pure Lanthanum \cite{petry} 
motivates an initial choice of the $\Lambda$ parameter of about one
\cite {andrews thesis}. Also, since we are not interested in the dynamics of
unstable systems we restrict our attention to $0 < \delta T < \delta T_{1}$.)
Then, we choose an initial displacement field that is both static
and supercritical; {\em e.g.}, for most of our dynamical studies we used the
supercritical state defined in Eq.~(\ref{eq:nucl test 1}) with
$\epsilon \approx 0.01$. Finally, we employ our numerical integration 
algorithm \cite {y/g} to follow the temporal evolution of our system.
We stress that the symmetry and/or profile of the initial state does not
qualitatively influence the interfacial dynamics observed, {\em viz.},
only the above--mentioned parameters influence the qualitative aspects of the
growth of the product phase.

For $\delta T$ just below the first--order transition temperature $T_{1}$, 
the profile in Fig. \ref{fig:two_kink} develops.  This type of dynamical 
evolution has been called a ``two--kink'' growth front --- also see
Fig. 1 of Ref. \cite{b/g}. Here we utilize the evolution shown
in Fig. 6a to develop an approximation for the growth interface found
in this temperature regime. This approximation provides us with an
excellent model from which an accurate theory of the instability 
of this interface can be fully developed.

Consider the representation of the displacement field shown in
Fig. \ref{fig:two kink approxn}. The similarity to the growth
interface of Fig. 6a for $x {> \atop \sim} 20$  is clear. 
For simplicity, we have shifted the ``back end" of the interface to 
be at $x = 0$, as well as reflecting the displacement field to be 
positive in the region of interest.
We do not possess an analytical expression for this interface profile,
and thus it is necessary to approximate its functional form.  This is
accomplished by ignoring the ``smoothing" imposed by the strain gradients;
this leads to the following piecewise continuous function for 
$u(x,t)$ \cite {smoothing comment}:
\begin{equation}
u(x,t) = \left\{ \begin{array}{llll}
                 0 && \mbox{, $x = 0$} \\
                 e_{m} x && \mbox{, $0 \leq x \leq v t$} \\
              (e_{m} + e_{2}) v t - e_{2} x && \mbox{, $v t \leq x \leq v_{s} t$
} \\
                 0 && \mbox{, $x \geq v_{s} t$}
                 \end{array}
                 \right.
\label{eq:piecewise_u}
\end{equation}
In this expression $e_{m}$ is the martensitic strain,
$e_{2}$ is the strain of the interface,
and the speeds $v$ and $v_{s}$ will be described and related below.

The physical constraint that the interface be coherent at $x = v_{s} t$
results in the following important relation between the speed $v$
and the speed of sound $v_{s}$ (see below), {\em viz}.,
\begin{equation}
v = \frac{e_{2}}{(e_{m} + e_{2})} v_{s}
\label{eq:growth_velocity}
\end{equation}
This one relationship, plus the analytical evaluation of $e_m$ (already
available in Eq.~(\ref{eq:emart})), 
$v$ and $v_s$, will allow for all of the following analysis.

An illustrative limiting case of Eq.~(\ref{eq:growth_velocity}) occurs when 
one considers $\Lambda \rightarrow 0$. This implies that the speed of sound 
$v_{s}$ becomes infinite, and thus Eq.~(\ref{eq:growth_velocity})
requires that $e_2 = 0$. This situation can 
be understood as the overdamped TDGLT limit of the two--kink solution \cite {b/g}.
It corresponds to a kink--type propagating solitary wave of amplitude $\pm e_{m}$
moving to the right/left at a speed $v$.
This corresponding analytical solution has already been found by Gordon 
\cite{gordon} and has the following form:
\begin{equation}
e(x,t) = \frac{\pm e_{m}}{(1 + {\rm exp}[(x - v t)/\ell])^{1/2}}
\label{eq:overdamped_e}
\end{equation}
where the speed $v$ and $\ell$ are uniquely determined by the undercooling
$\delta T$.

Unfortunately, this solution is entirely unphysical. (This is an important
point since discussions in the literature have used such a state as a model 
of martensitic domain wall propagation under the name of the Eshelby model.)
The finite propagation time of the elastic field for the physically
relevant scenario of $\Lambda \neq 0$ immediately suggests that 
such motion must have an unbounded kinetic energy \cite{b/g}.
For example, if one calculates the kinetic energy $K$ of
Eq.~(\ref{eq:overdamped_e}) using Eq.~(\ref{eq:kinetic energy})
for a system of length $L$ one finds that $K \propto L$, a result
that simply reflects the fact that for this propagating interface
the entire system is moving! Clearly, this high energy state is
never going to be selected as the path through which the system
approaches its steady state.
Thus, the unphysical result contained in Eq.~(\ref{eq:overdamped_e}) serves
to emphasize the importance of including the inertia of the displacement field 
in any accounting of the interfacial dynamics of martensitic
growth fronts \cite {b/g}. With this in mind, we return to the
physically relevant $\Lambda \not= 0$ problem. 

We wish to describe the interface displayed in Fig.~6a,
and to this end we note that the complete analytical characterization of 
this propagating interface requires
the evaluation of the speed of sound $v_{s}$, possibly nonlinear, and the
forward kink amplitude $e_{2}$.
Note that in the moving interface portion of Fig. 7, $v_s$ characterized
the motion in the small strain ($e_2$) region, and for this reason we
evaluate $v_s$ by determining the nonlinear speed of sound.
In a linear theory, the elastic field propagates at a characteristic velocity
determined by the undercooling $\delta T$ and mass density $\Lambda$, {\em viz}.,
\begin{equation}
v_{s}^{2} = \frac{\delta T}{\Lambda}
\label{eq:linear_vs}
\end{equation}
We require an expression for the nonlinear speed of sound, and to this end we
consider the ``nonlinear corrections'' to Eq.~(\ref{eq:linear_vs}), {\em viz.},
\begin{equation}
v_{s}^{2} = \frac{\delta T + a e^{2} + b e^{4} + \cdot \cdot \cdot}{\Lambda}
\label{eq:corrections_vs}
\end{equation}
where $a, b, ...$, are temperature independent constants and, by symmetry, we
have only included the even powers of the strain $e$.   Then, we turn to the
local, nondispersive equation of motion for the strain field which can be
written as
\begin{equation}
\partial_{t}^{2} e = \frac{1}{\Lambda} \partial_{x}^{2} [(\delta T - e^{2} + e^{4}) e]~.
\label{eq:local_eom_e}
\end{equation}
Locally, the term $(\delta T - e^{2} + e^{4})$, 
approximately corresponds to the nonlinear
force constant in the parent phase, which we will denote as $\tilde{c}$.
Comparing Eq.~(\ref{eq:corrections_vs}) 
and Eq.~(\ref{eq:local_eom_e}) we see that if $a = -1$ and $b = 1$, 
\begin{equation}
v_{s}^{2} = \frac{\tilde{c}}{\Lambda} \approx \frac{1}{\Lambda} (\delta T - e^{2} + e^{4})
\label{eq:nonlinear_vs}
\end{equation}
represents the $4$--th order nonlinear corrections to the speed of sound of 
an elastic wave of amplitude $e$.  In
particular, for the nonlinear speed of sound $v_{s}$ of the 
strain amplitude $e_{2}$ interface, we shall use
\begin{equation}
v_{s} \approx \sqrt{\frac{1}{\Lambda} (\delta T - e_{2}^{2} + e_{2}^{4})}~.
\label{vs_forward_kink}
\end{equation}

The last quantity that we need to determine is the amplitude $e_{2}$ of the
forward kink.  A physically motivated method for obtaining this quantity
relies on energy considerations.  According to the Lagrangian dynamical 
formalism \cite{Landau}, the time rate of change of the energy $E$ of the
system is related to the dissipation function $R$ by 
\begin{equation}
\partial_{t} E = - 2 R~.
\label{eq:energy_balance}
\end{equation}
Thus, in order for us to derive an expression for $e_{2}$ we need construct
the total energy $E$ of the system and the dissipation function $R$.

The energy of the system is readily calculated by using our approximation
for the displacement field $u(x,t)$, {\em viz}., Eq.~(\ref{eq:piecewise_u}), and
the following expression for the total energy density $\cal E$:
\begin{equation}
{\cal E} = \frac{1}{2} \Lambda (\partial_{t} u)^{2} + (~\frac{1}{2} \delta T e^
{2} -
\frac{1}{4} e^{4} + \frac{1}{6} e^{6}~)~. 
\label{eq:mech_energy_density}
\end{equation}
The total energy is then simply the piecewise spatial integration of the 
above equation
over the regions $I_{1} = [0, v t]$, and $I_{2} = [v t, v_{s} t]$.  The 
result of such a calculation yields an expression for E, {\em viz}.,
\begin{equation}
E = \frac{e_{m} e_{2} v_{s}}{(e_{m} + e_{2})} [ \delta T e_{
2} - \frac{3}{4} e_{2}^{3} + \frac{2}{3} e_{2}^{5} + \frac{1}{2} \delta T e_{m}
- \frac{1}{4} e_{m}^{3} + \frac{1}{6} e_{m}^{5}]~t~,
\label{eq:E_dot}
\end{equation}
and $\partial_{t} E$ is given by a single time derivative of the above equation.

Using the same procedure as in \cite{b/g}, we approximately solve for the
dissipation function through the use of the
analytical expression for the overdamped kink, {\em viz}., 
Eq.~(\ref{eq:overdamped_e}).  At long times, the dissipation function can
be written as \cite{b/g},
\begin{equation}
R \approx \frac{e_{m}^{2}}{16 \ell} v^2
\label{eq:approx_R}
\end{equation}
where $v$ is the growth speed and $\ell$ is a measure of the interfacial
width ($\ell \sim 1$ domain wall width).  The energy balance condition, 
Eq.~(\ref{eq:energy_balance}), can now be restated in terms of
the only, {\em a priori}, unknown quantity, namely $e_{2}$, {\em viz}.,
\begin{equation}
\frac{e_{m} e_{2} v_{s}}{(e_{m} + e_{2})} [ \delta T e_{2} - \frac{3}{4} e_{2}^{
3} + \frac{2}{3} e_{2}^{5} + \frac{1}{2} \delta T e_{m}
- \frac{1}{4} e_{m}^{3} + \frac{1}{6} e_{m}^{5}] - \frac{e_{m}^{2}}{8 \ell} v^2
 = 0~,
\label{eq:expr_e2}
\end{equation}
where it should be recalled that $v_{s}$ and $v$ can both be expressed
in terms of $e_{2}$.  We have tested this calculation by
numerically solving for $e_{2}$ in Eq.~(\ref{eq:expr_e2}) and then
using this in Eq.~(\ref{eq:growth_velocity}) to obtain a value for $v$.  We have
found that for the parameters
$E_{W}/E_{B} = 1$, $\Lambda = 1$, $\ell = \sqrt{3}/(2 e_{m}^{2})$,
$v = 0.035$, which
compares superbly with the value found from the numerical integration of
the full dynamical equations, {\em viz}., $v_{num} = 0.033$.  
Similar favourable comparisons are found
for other choices of $\delta T$ and $\Lambda$ \cite {length v}.  
This excellent agreement between the analytical and numerical values 
for the growth speed $v$ clearly justifies the approximation that we
have used for $u(x,t)$ in Eq.~(\ref{eq:piecewise_u}).

One may relate this speed to that in the proper, 1st--order, ferroelastic
transition undergone by Lanthanum.  One predicts that
the two--kink interfacial speed would be of the order of 10 \%
(or less) of the smallest speed of sound, namely about 80 $m/s$.

We have found that this interfacial motion does not persist --- instead,
as the temperature of the system is lowered 
a second type of propagating growth front develops. 
Figure 8 illustrates a typical solution for both the 
displacement and strain fields. The strain figure makes clear
the fascinating physics that Bales and one of us \cite {b/g} observed.
This growth front has been called ``dynamical twin formation'' 
\cite{b/g} because the moving interface separating product from parent 
phase leaves behind an alternating structure 
consisting of both doubly degenerate low--temperature variants --- 
these are so--called martensitic twins, and in Fig. 8c we display
the steady--state profile of a short twinned crystal that results
from the dynamics shown in Figs. 8a/b.  

Below we shall describe the origin of this interfacial motion.
However, here we make one straightforward observation contrasting this type of 
interfacial motion with that of Fig.~\ref{fig:two_kink}.  Figure 
\ref{fig:twin_KE} shows a snapshot of the kinetic energy density, 
$\tilde K(x,t)$, for the dynamical twinning solution at time $t = 360$ scaled 
time units,  and illustrates the dramatic localization of the kinetic energy 
density around the interface ($x \approx 75$) of the transformation front.
This is clearly different from the entirely delocalized kinetic energy
density found throughout the $e_2$ interface shown in Fig.~\ref{fig:kink_KE}
(also $t = 360$ scaled time units) --- note the very different scales of
these two figures.

Both growth profiles can be understood as arising from the finite
propagation time of the elastic field found for $\Lambda \neq 0$.
Recall that in our approach sound waves always propagate at sufficiently
long wavelengths.  It then follows that the displacement field far
from the interface must remain fixed, and thus there is an induced stress
in the immediate vicinity of the growth front; that is, the parent phase
is bent locally at the interface.
The magnitude of the stress induced in the parent
phase depends strongly on the material
parameters, $\Lambda$ and $\delta T$, through Eq. (\ref{eq:growth_velocity}).
So, as the growth speed approaches the speed of sound, 
the $e_2$ strain increases and will eventually
exceed the spinodal strain \cite{spinodal comment}
determined by 
$\partial^{2} f_{L}(e)/\partial e^{2} = 0$, {\em viz}., 
\begin{equation}
e_{ss} = \sqrt{0.1 (3 - \sqrt{9 - 20 \delta T})}~. 
\label{eq:espin}
\end{equation}
When this occurs the (local) interface becomes unstable and a dynamical
phase transition arises producing a new type of interfacial motion,
{\em viz.} that shown in Fig. 8a/b.
Summarizing, we see that the dynamical twinning shown in Fig. 8a/b
is a consequence of the local instability of the $e_2$--strain interface of Fig. 6 a/b
separating product from parent phase. It is worthwhile
to re--emphasize that although the $t=0$ initial states discussed in
this paper were twinned nuclei, the
dynamical evolutions illustrated in Fig. 6 and Fig. 8 are {\em independent}
of the initial displacement field ({\em e.g.}, see also Fig. 1 and 2 of \cite{b/g}
where different initial conditions were used).

Based on the above, one can expect that in ($\Lambda$,$\delta T$) phase 
space, a twinning/no--twinning phase diagram may be constructed. 
By setting $e_{2} = e_{ss}$ in Eq. (\ref{eq:expr_e2}) and substituting
Eq. (\ref{eq:growth_velocity}) in for $v$, we can solve for the critical
value 
$\Lambda_c (\delta T)$ \cite {PRL_correction}, {\em viz}.,
\begin{equation}
\Lambda_c^{1/2} = \frac{- e_{m}^{3} e_{ss} (\delta T - e_{ss}^{2} + e_{ss}^{4})^{1
/2}}{4 \sqrt{3} (e_{m} + e_{ss}) [\delta T e_{ss} - \frac{3}{4} e_{ss}^3 + \frac
{2}{3} e_{ss}^5 + \frac{1}{2} \delta T e_{m} - \frac{1}{4} e_{m}^{3} + \frac{1}{
6} e_{m}^{5}]}~.
\label{eq:twinning_boundary}
\end{equation}
where it is to be recalled that both $e_{m}$ and $e_{ss}$ are explicit 
functions of $\delta T$ through Eqs. (\ref{eq:emart},\ref{eq:espin});
viewed in ($\Lambda$,$\delta T$) phase space, this trajectory defines the
critical value of $\Lambda$ at which twinning occurs.  This curve is plotted in
Fig. \ref{fig:twin_boundary} along with the results from the numerics.
The divergence of Eq. (\ref{eq:twinning_boundary}) when 
$\delta T \cong 0.136$ defines the range of undercoolings for which
twinning will not occur for {\em any} value of $\Lambda$.  The superb
agreement of Eq. (\ref{eq:twinning_boundary}) with that of our numerics clearly
justifies our theory of the instability of the growth interface.  
\section{Summary and Discussion:}
\label{sec:sum_discuss}

We have presented an exhaustive study of the dynamical nucleation and
growth in a \hfill\break one--dimensional version of a
first--order, shear elastic martensitic phase transformation.
Our approach involves treating nonlinear elastic models of these
transitions as nonconvective, hydrodynamic systems.
We have been able to determine for the first time that the bulk critical nuclei
of the system are twinned structures.  We have obtained approximate
analytical expressions for these structures, and a comparison of 
Fig. 3b and Fig. 5 shows that our results are
in superb agreement with the exact numerical solutions.  In addition, we
have exactly solved for the critical nucleus of a finite, but large system
with open boundary conditions, {\em viz}., 
Eqs. (\ref{eq:bulk zeroforce},\ref{eq:bc1 zeroforce},\ref{eq:bc2 zeroforce}).
We have shown that, for this case, the critical nucleus is
a localized surface state with (almost) exactly $1/4$ the energy of the
bulk critical nucleus.  We conclude that surface nucleation 
is the dominant route for the formation of martensite,
at least in the absence of impurities.

A full treatment of the martensitic growth dynamics associated with 
supercritical
nuclei has also been presented.  A phase diagram connecting the rescaled
mass coefficient $\Lambda$ and the undercooling $\delta T$ has been derived.
We have shown that the transformation fronts propagate through the system
with a speed given by Eq. (\ref{eq:growth_velocity}). 

The extension of our work to higher
dimensions is in principle relatively straightforward. In fact,
a large amount of numerical work examining a triangular--to--oblique
transition has already been completed \cite{andrews thesis}.  
The formal ingredients are the same as
the one--dimensional model, except that the numerical
analysis becomes {\em much} more difficult --- the restrictions to higher
dimensional systems are essentially technical ones.  New integration
software will be needed in order to fully investigate the dynamical 
evolution of such systems.  Analytical work will be, obviously,
greatly complicated.

Although extending our one--dimensional phenomenology to completely general
2D and 3D systems can pose substantial technical problems, we have had 
much success in looking at higher dimensional models with special symmetries.
In particular, we have been able to use the {\em same} methodology as in
our 1D studies to investigate change of volume first--order transitions in a 
$d$--dimensional, nonlinear, nonlocal elastic system ($d = 1, 2, ..., \infty$).
In these models, the system is reduced down to $1 + 1$ dimensions for which
the variant method of lines and Lagrangian formalism presented in this 
paper are directly applicable in {\em all} $d$--dimensions. For these systems
we find that only surface nuclei exist, and that all dynamical evolutions
of the system to steady state pass through such nuclei --- these results
will be presented elsewhere.

\acknowledgments

One of us (RJG) wishes to thank the many colleagues who have worked
with him on this problem over the years: Jim Krumhansl, Baruch Horovitz,
Gerhard Barsch, Alan Bishop, Jamie Morris, Steve Bales, and Andrew Reid.
We also wish to thank Bob Kohn, John Ball and Dick James for clarifying 
comments on competing theories of these transitions.
We appreciate the helpful comments of Ken Vos concerning the solution
discussed in Appendix B.
This work was supported by the NSERC of Canada, and the
Advisory Research Committee of Queen's University. 

\newpage

\appendix
\section{Dimensional Analysis:}
\label{sect:diman}

Following the notation of Ref. \cite {Bluman}, let $u$ be the displacement 
at any position in space, and assume that
\begin{equation}
u = f(W_{1},W_{2}, ..., W_{10})     
\label{eq:straindim}
\end{equation}
where
\mbox{}\\
$W_{1} = A$, some positive constant of dimension $[A] = \frac{m \ell}{\theta \tau^{2}}$,
\mbox{}\\
$W_{2} = B$, some positive constant of dimension $[B] = \frac{m \ell}{\tau^{2}}$,
\mbox{}\\
$W_{3} = C$, some positive constant of dimension $[C] = [B]$,
\mbox{}\\
$W_{4} = D$, some positive constant of dimension $[D] = \frac{m \ell^{3}}{\tau^{2}}$,
\mbox{}\\
$W_{5} = \gamma$, the sound wave viscosity of dimension
$[\gamma] = \frac{m \ell}{\tau}$,
\mbox{}\\
$W_{6} = \rho$, the linear mass density of dimension
$[\rho] = \frac{m}{\ell}$,
\mbox{}\\
$W_{7} = \delta T$, the undercooling of dimension
$[\delta T] = \theta$,
\mbox{}\\
$W_{8} = L$, the system size of dimension $[L] = \ell$
\mbox{}\\
$W_{9} = x$, the distance along the bar of dimension $[x] = \ell$,
\mbox{}\\
$W_{10} = t$, the elapsed time after an initial strain is applied of
dimension $[t] = \tau$.

For this analysis, we have used {\it dynamical units} as our fundamental basis,
{\em viz}., 
\mbox{}\\ 
$L_{1} = \ell$ (length)
\mbox{}\\
$L_{2} = m$ (mass)
\mbox{}\\
$L_{3} = \tau$ (time)
\mbox{}\\
$L_{4} = \theta$ (temperature)
\mbox{}\\
The corresponding dimension matrix is then simply:
\begin{equation}
{\bf B} = \left[ \begin{array}{rrrrrrrrrr}
          1&1&1&3&1&-1&\;0&\;1&\;1&\;0\\
          1 &1 &1&1&1&1&\;0&\;0&\;0&\;0\\
         -2&-2&-2&-2&-1&0&\;0&\;0&\;0&\;1\\
          -1&0&0&0&0&0&\;1&\;0&\;0&\;0\\
\end{array} \right]
\end{equation}

This matrix has a rank $r({\bf B}) = 4$, whence by the Buckingham Pi--Theorem 
\cite {Bluman}, the number of measurable dimensionless quantities, 
under our choice of fundamental units, is
$k = n - r(B) = 10 - 4 = 6$, where $n$ is the total number of independent
variables and constants appearing in the system.
It also follows from the Buckingham Pi--Theorem that our system can be
re--expressed in dimensionless form where 
$\pi = \tilde{u}(\Omega)$ is a dimensionless
dependent variable and $\Omega =$ \{$\pi_{1}, \pi_{2}, ... , \pi_{k}$\} 
are dimensionless
{\it independent} variables and dimensionless constants.  
The exact form of these
$k$ dimensionless quantities can be found by solving for the {\it null space}
of the dimension matrix ${\bf B}$.  The dimensionless displacement field
$\tilde{u}(\Omega)$ is found by solving the linear system 
${\bf B y} = - {\bf a}$, where {\bf a} is the dimension vector of $u$.
Tedious algebra reveals that $\Omega$ and $\pi$ can be written as
\begin{equation}
\pi_{1} = \frac{B}{\gamma}~t,~\pi_{2} = \left[\frac{B}{D}\right]^{1/2}x,~
\pi_{3} = \left[\frac{B}{D}\right]^{1/2} L,~
\pi_{4} = \frac{A}{B}~\delta T,~                
\pi_{5} = \frac{\rho D}{\gamma^{2}},~
\pi_{6} = \left[\frac{C}{B}\right]^{1/2}~,
\end{equation}
and
\begin{equation}
\pi = \left[\frac{B}{D}\right]^{1/2}~u~.
\label{eq:rescaled u}
\end{equation}
If we identify $\pi_{5} = \Lambda$, a rescaled mass coefficient, and rescale
$\pi_{1}, ..., \pi_{4}$ using $\pi_{6}$, we arrive at the
following dimensionless, rescaled independent variables and constants,
\begin{equation}
\tilde{t} =  \frac{\pi_{1}}{\pi_{6}^{2}} = \frac{B^{2}}{\gamma C}~t~,~~~~
\tilde{x} =  \frac{\pi_{2}}{\pi_{6}} = \frac{B}{\sqrt{C D}}~x,~ 
\label{reduced ind}
\end{equation}
\begin{equation}
\tilde{L} =  \frac{\pi_{3}}{\pi_{6}} = \frac{B}{\sqrt{C D}}~L,~~~~ 
\delta \tilde{T} = \pi_{6}^{2} \pi_{4} =  \frac{A C}{B^{2}}~\delta T,~~~~
\frac{\rho D}{\gamma^{2}} = \Lambda~.
\label{eq:reduced omega}
\end{equation}

These scaled quantities, along with Eq.~(\ref{eq:rescaled u}), can now 
be used in Eqs.~(\ref{eq:bulk equation}), 
(\ref{eq:bc1}) and (\ref{eq:bc2}), to obtain 
a complete two--parameter
model which depends only on a rescaled undercooling ($\delta \tilde{T}$) and
a rescaled mass coefficient ($\Lambda$), {\em viz}.,
\begin{equation}
\Lambda (\partial^{2}_{\tilde{t}} \tilde{u}) = (\partial^{2}_{\tilde{x}} \tilde{u}) [~\delta \tilde{T} - 3 (\partial_{\tilde{x}
} \tilde{u})^{2}
+ 5 (\partial_{\tilde{x}} \tilde{u})^4~] - (\partial^{4}_{\tilde{x}} \tilde{u}) + (\partial_{\tilde{t}} \partial ^{2}_{\tilde{x}
} \tilde{u})~,
\label{eq:appendix eom}
\end{equation}
with boundary conditions at $\tilde{x} = \pm \frac{\tilde{L}}{2}$,
\begin{equation}
\delta \tilde{T} (\partial_{\tilde{x}} \tilde{u}) - (\partial_{\tilde{x}} \tilde{u})^{3} + (\partial_{\tilde{x}} \tilde{u})^{5} +
(\partial_{\tilde{x}} \partial_{\tilde{t}} \tilde{u}) - (\partial^{3}_{\tilde{x}} \tilde{u}) = 0,
\label{eq:appendixbc1}
\end{equation}
\begin{equation}
(\partial^{2}_{\tilde{x}} \tilde{u}) = 0~.
\label{eq:appendixbc2}
\end{equation}
\section{Exact Solutions for an Infinite Domain:}

In terms of the shear strain, the bulk, zero--force equation that we wish
to solve for is given by
\begin{equation}
\partial_{x}^{2} e - (~\delta T e - e^{3} + e^{5}~) = 0
\label{eq:strain_bulk_zeroforce}
\end{equation}
where the boundary conditions are $\partial_{x}^{j} e(x) = 0$ at 
$x = \pm \infty$ and $j = 0, 1, ..., 4$.
\mbox{}\\
If we multiply Eq.~(\ref{eq:strain_bulk_zeroforce}) by $\partial_{x} e$ and
integrate, we obtain
\begin{equation}
(\partial_{x} e)^{2} - (~\delta T e^{2} - \frac{1}{2} e^{4} + \frac{1}{3} e^{6}~) = c       
\label{eq:integration}
\end{equation}
where $c$ is an integration constant.  From the constraint that all derivatives
of the strain must vanish at $\pm \infty$, it follows that $c = 0$.
\mbox{}\\
Let $e = \pm 1 / \sqrt{w}$, so that we are now looking for singular solutions
of $w$.  Note that this transformation is well defined because $e(x)$ is
bounded for all $x \in (-\infty, \infty)$.  Equation~(\ref{eq:integration}),
with $c = 0$, then becomes:  
\begin{equation}
(\partial_{x} w)^{2} - (~4 \delta T w^{2} - 2 w + \frac{4}{3}~) = 0
\label{eq:sub1}
\end{equation}
\mbox{}\\
Re--writing this as an integral over $w$ and $x$, we have,
\begin{equation}
\int~\frac{dw}{\sqrt{4 \delta T w^{2} - 2 w + \frac{4}{3}}} = \int~dx
\label{eq:integral}
\end{equation}
Both the left and right hand sides of Eq.~(\ref{eq:integral}) can be 
evaluated exactly to give the following expression for $w$ and $x$
\begin{equation}
\frac{\ln (-1 + 4 \delta T w + \sqrt{\frac{8 \delta T}{3}} \sqrt{2 - 3 w + 6 \delta T w^{2}})}{2 \sqrt{\delta T}} = x - x_{0}
\label{eq:w_and_x}
\end{equation}
where $x_{0}$ is an integration constant.
The inversion of this equation will yield $w$ as a function of $x$, {\em viz}.,
\begin{equation}
w(x) = \frac{3 - 16 \delta T + 6~{\rm exp}[2 \sqrt{\delta T} (x - x_{0})] +
3~{\rm exp}[4 \sqrt{\delta T} (x - x_{0})]}{24 \delta T {\rm exp}[2 \sqrt{\delta T} (x - x_{0})]}
\label{eq:eqn_z}
\end{equation}
Solutions to 
Eq.~(\ref{eq:strain_bulk_zeroforce}) are finally obtained by recalling that
$e = \pm 1/\sqrt{w}$, whence,
\begin{equation}
e_{0}(x-x_{0}) = \pm 2 \sqrt{6 \delta T} \sqrt{\frac{{\rm exp}[2 \sqrt{\delta T
} (x - x_{0})]}{3 - 16 \delta T + 6~{\rm exp}[2 \sqrt{\delta T} (x - x_{0})] +
3~{\rm exp}[4 \sqrt{\delta T} (x - x_{0})]}}
\label{eq:appdx strain}
\end{equation}
A direct substitution of Eq. (\ref{eq:appdx strain}) into
Eq. (17) confirms that they are indeed zero--force solutions.


\begin{figure}
\caption{A plot of the local elastic free--energy density, $f_L(e)$, 
of Eq. (1) for $A = B = C = 1$ (corresponding to the scaled units used
in our nucleation and growth studies) and $\delta T = 1/6$. The barrier
height is denoted by $E_{B}$ and the well depth by $E_{W}$.}
\label{fig:potentialfig}
\end{figure}

\begin{figure}
\caption{The classical nucleus (long--dash line), corresponding 
to $n \rightarrow \infty$ in Eq. (21), and a 
Gaussian fluctuation (dot--dashed line) corresponding to $n = 1$.  
Again referring to Eq. (21), for these figures $d = 10$ and $u_{0} = 1$.}
\label{fig:limiting_nuclei}
\end{figure}

\begin{figure}
\caption{(a) The bulk critical nucleus found from solutions to Eqs.~(17,18,19) 
for $E_{W}/E_{B} = 0.2, 1/6$, and $5$ --- the tallest curve corresponds
to the smallest ratio of $E_{W}/E_{B}$. \hfill\break
(b) The bulk critical nucleus in 
strain space for the ratio $E_{W}/E_{B} = 1.0$.}
\label{fig:critical_nucleus} 
\end{figure}

\begin{figure}
\caption{A plot of Eq. (25) for $E_{W}/E_{B} = 1.0$.  
This curve has been shifted so that $x = 0$ corresponds to the $x_{max}$ 
given in Eq. (26).}
\label{fig:soliton} 
\end{figure}

\begin{figure}
\caption{A plot of the analytical approximation to the critical nucleus
for $E_{W}/E_{B} = 1.0$ based on Eq. (27) using the procedure described
in the text for evaluating $x_{0}$. This strain profile should be compared
with that shown in Fig. 3b, the (numerically determined) exact
critical nucleus for this temperature. }
\label{fig:approx. strain nuclei}
\end{figure}

\begin{figure}
\caption{(a) Evolution of the displacement field for a supercritical
nucleus ($\epsilon = +0.01$ in Eq. (23)) for $E_{W}/E_{B} = 1.0$. The dashed line
is the (static) initial displacement field, and the 
solid lines show how the
system progresses at the times $t = 0$, $180$, $220$, $260$ and $300.$
(b) Evolution of the strain field for the supercritical nucleus studied 
in (a); $e_{m}$ is the martensite strain given by Eq. (16).
The dashed line is the (static) initial strain field, and the 
solid lines show how the system progesses at the times $t = 0$, 
$400$, $500$ and $600$.
(c) The steady--state $t \rightarrow \infty$
profiles of the displacement field (solid line) and the strain field 
(dashed line). In the context of our model, this steady state corresponds 
to a single twin or ``bicrystal''.}
\label{fig:two_kink}
\end{figure}

\begin{figure}
\caption{An idealized representation of the two--kink growth interface.
For the $t=0$ state being the dashed line, the solid line shows
the interface at some later time $t$.  $v$ is the interfacial growth speed
and $v_{s}$ is the nonlinear speed of sound, a quantity that we have attempted
to evaluate in Eq. (38).  The
exclusion of the nonlocal term, {\em viz}., $\partial^{4}_{x} u$, simplifies the
problem to finding only a $C^{1}$ function for $u(x,t)$.}
\label{fig:two kink approxn}
\end{figure} 

\begin{figure}
\caption{(a) Evolution of the displacement field for a supercritical 
nucleus ($\epsilon = +0.01$ in Eq. (23)) for $E_{W}/E_{B} = 6.5$.  The
dot--dashed line is the (static) initial displacement field, and the 
long--dashed line corresponds to a time of $t=360$, and the solid
line represents a time of $t=720$.
(b) Evolution of the strain field for the supercritical nucleus studied
in (a); $e_{m}$ is the martensite strain given by Eq. (16).
(c) The steady--state $t \rightarrow \infty$
profiles of the displacement field (solid line) and the strain field 
(dashed line).  In the context of our model, this steady state corresponds to a
polytwinned crystal. }
\label{fig:twinning}
\end{figure}

\begin{figure}
\caption{The kinetic energy density for the dynamical twinning growth 
front at $t = 360$ for $E_{W}/E_{B} = 6.5$.
For $x {> \atop \sim} 75$, the
kinetic energy density is almost completely localized (see text).}
\label{fig:twin_KE}
\end{figure}

\begin{figure}
\caption{The kinetic energy density for the two--kink growth front at
$t = 360$ for $E_{W}/E_{B} = 1.0$.
For $x {> \atop \sim} 75$, the kinetic energy density
is spread out over the $e_{2}$ growth interface.}
\label{fig:kink_KE}
\end{figure}

\begin{figure}
\caption{The renormalized mass density, $\Lambda$, as a function of
the undercooling $\delta T$.  The region labeled $(I)$ corresponds to those
values of $(\Lambda, \delta T)$ for which twinning will occur.  Region $(II)$
will only result in the two--kink profile shown in Fig. 6a/b. 
$\delta T_{1} = 0.1875$ is the transition temperature in scaled units.
The solid circles are the numerical data of Ref. [4].}
\label{fig:twin_boundary}
\end{figure}


\begin{references}
\bibitem{smoothing comment}
This idealization can be thought of as a {\em local}
approximation to the two--kink solution, {\em viz}., the nonlocal term 
$\partial^{4}_{x} u$ in the equation of motion Eq.~(\ref{eq:bulk equation}),
which is responsible for the ``smoothing'' of $u(x,t)$, is neglected.
Formally, by ignoring the nonlocal term, we only require
a $C^{1}$ differentiable function to approximate $u(x,t)$.
\bibitem{length v}
We have also verified that
the growth speed is indeed independent of $L$.  This was accomplished by 
changing the length of the system ({\em e.g.}, doubling $L$) and noting that $v$
was invariant with respect to this change.
\bibitem{spinodal comment}
The spinodal strain generates a $(e, \delta T)$
phase diagram that approximates the 
unstable coexistence of the two--kink and dynamical twinning growth front
solutions.  If $e_{2} < e_{ss}$, dynamical twin formation will never
develop.
\end{references}
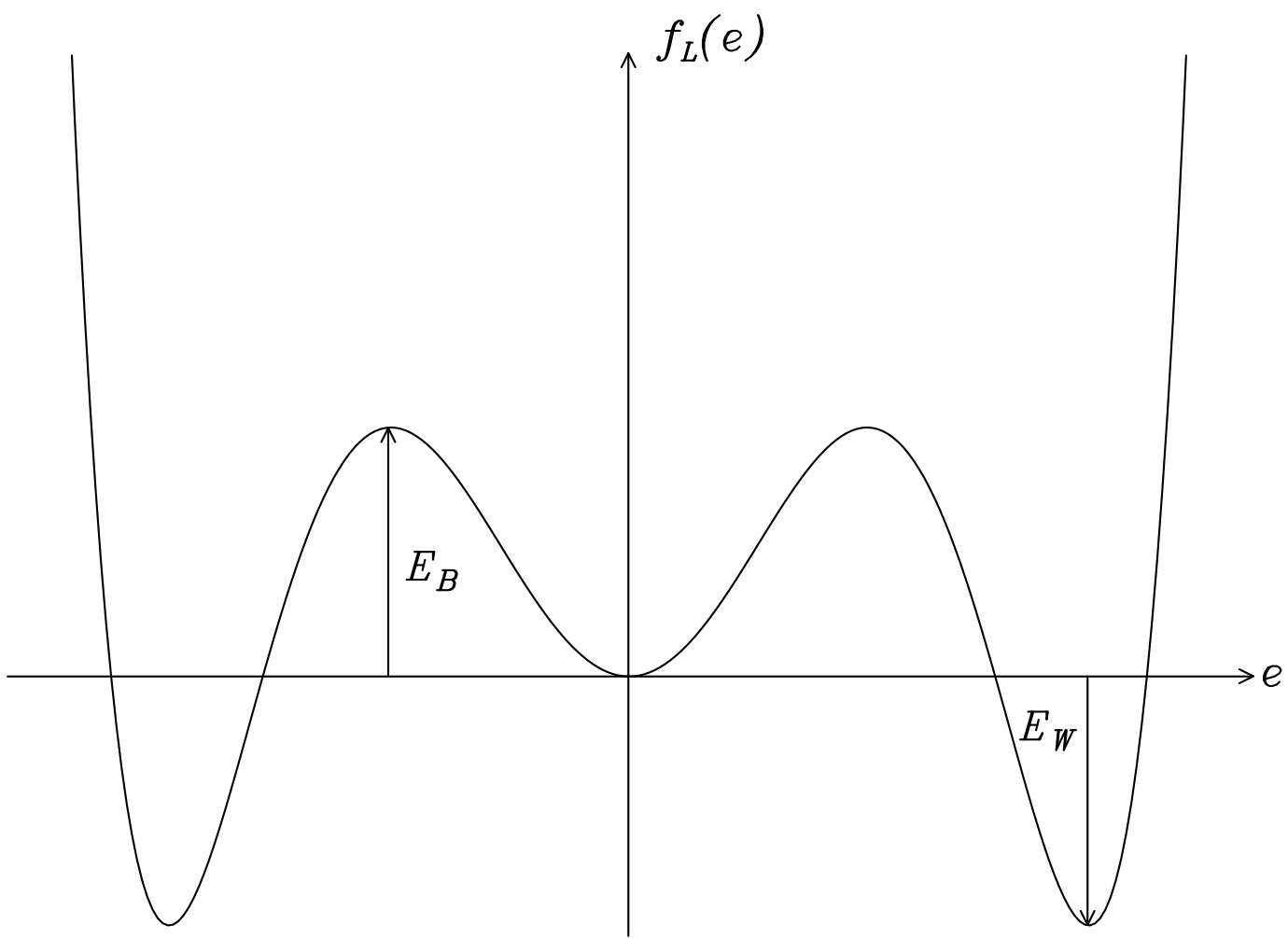
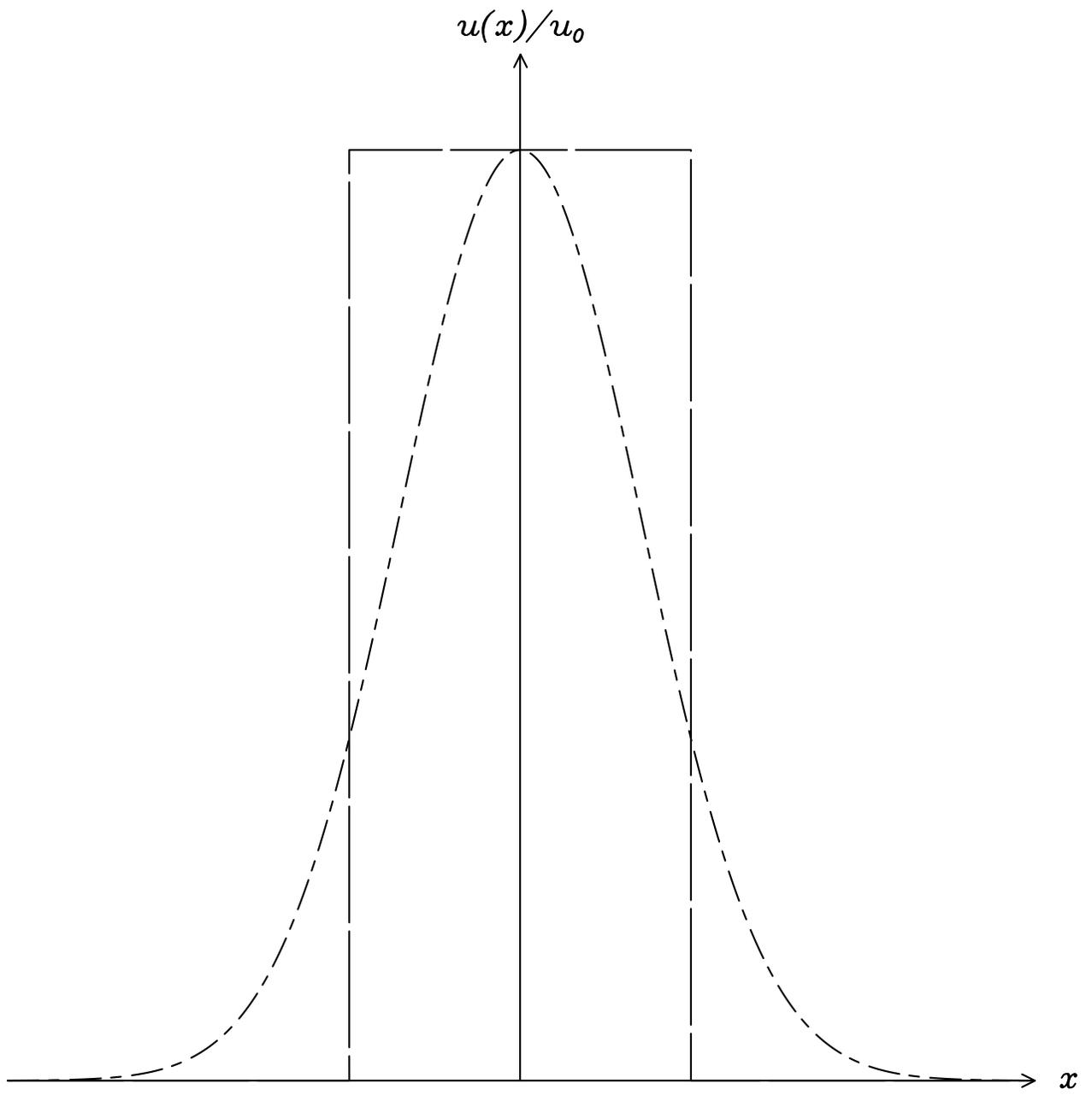
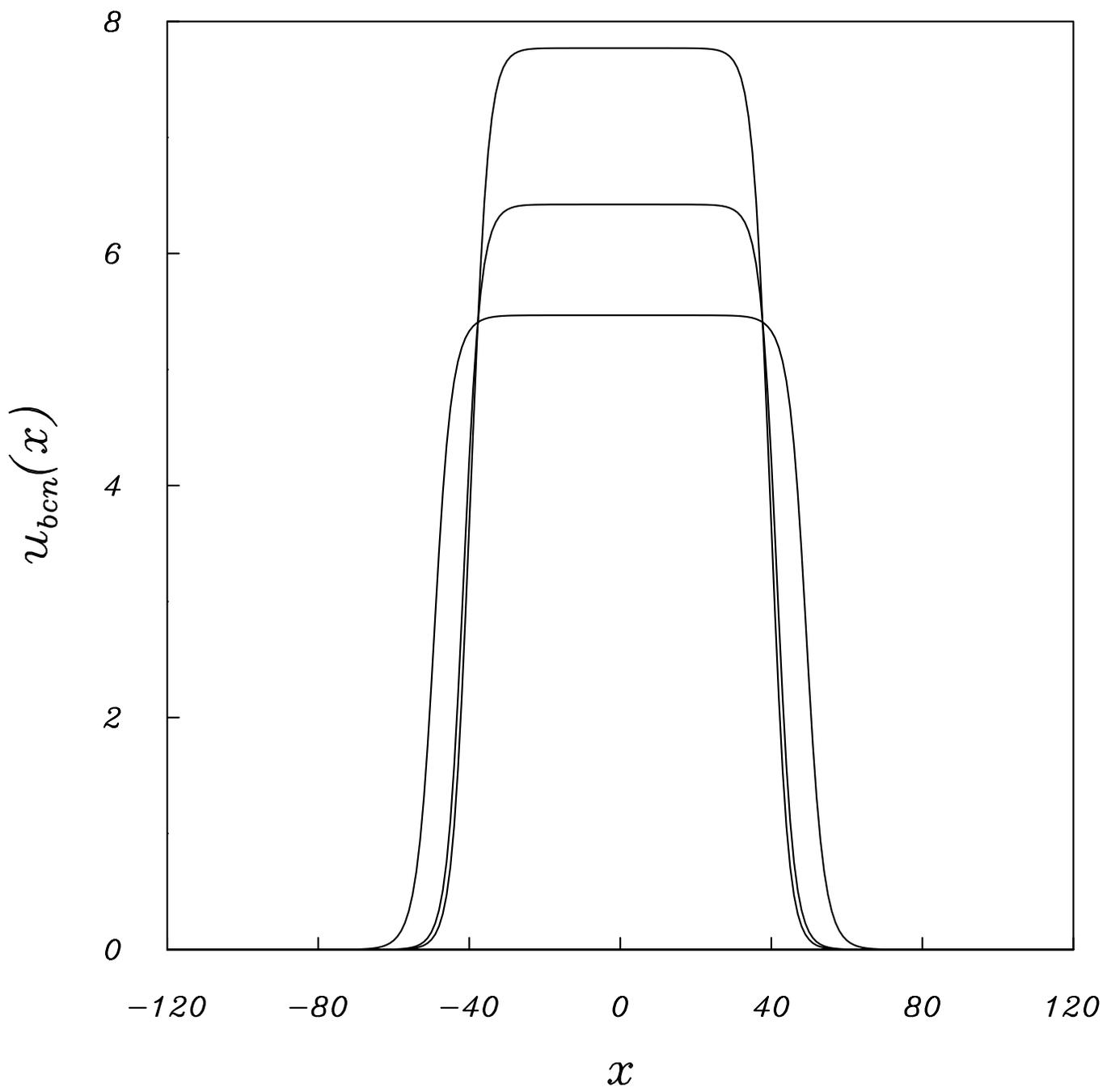
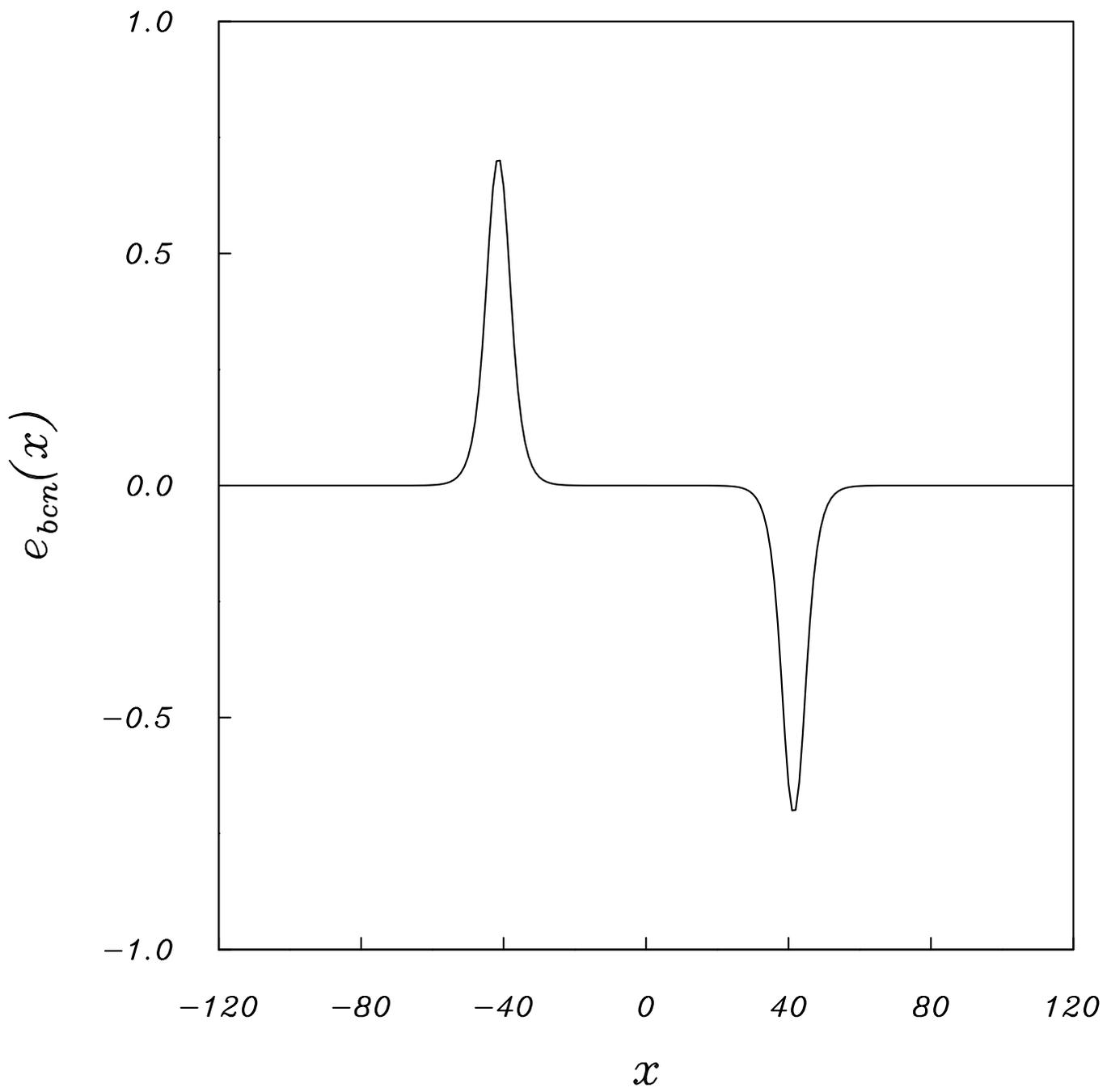
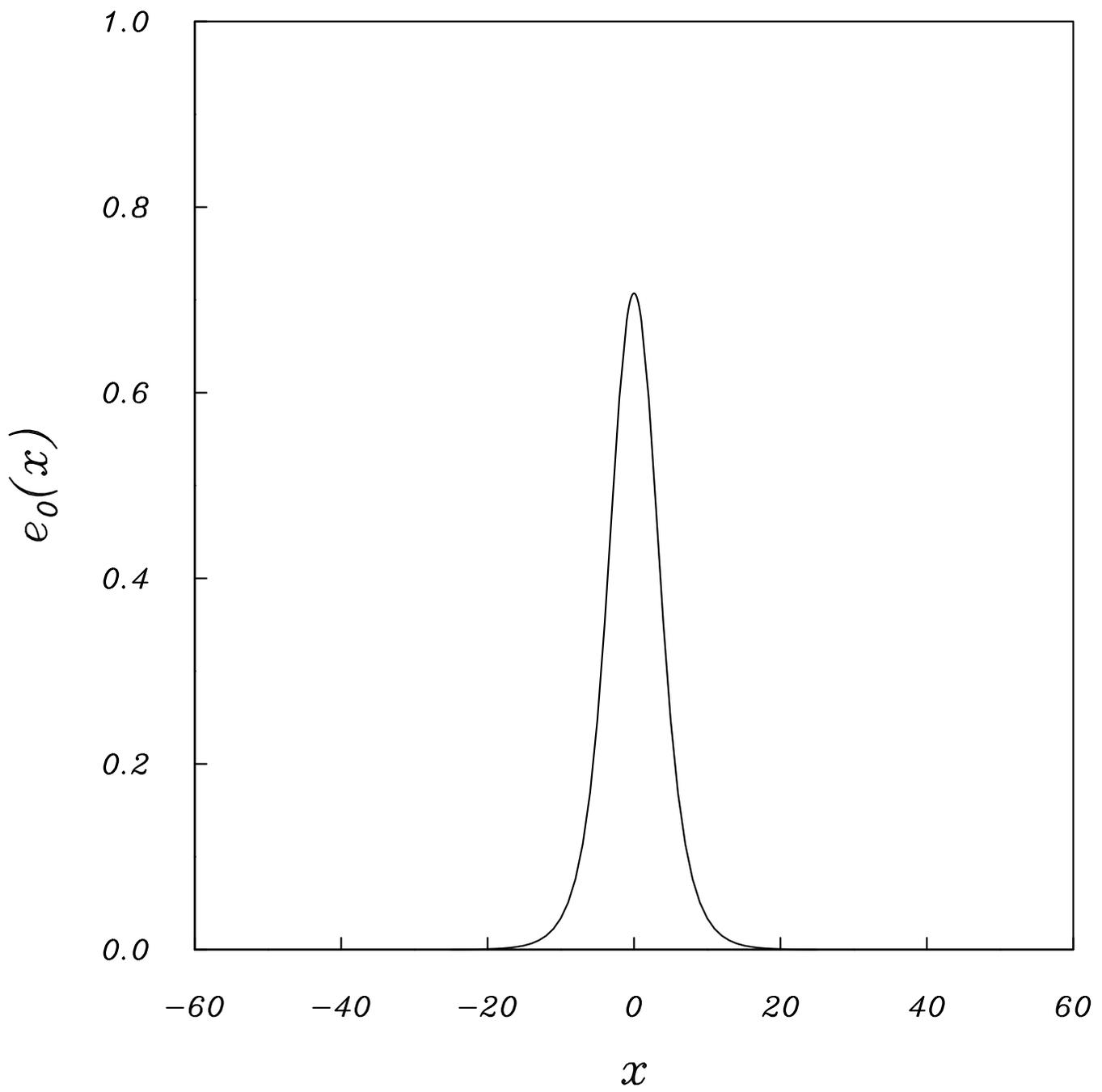
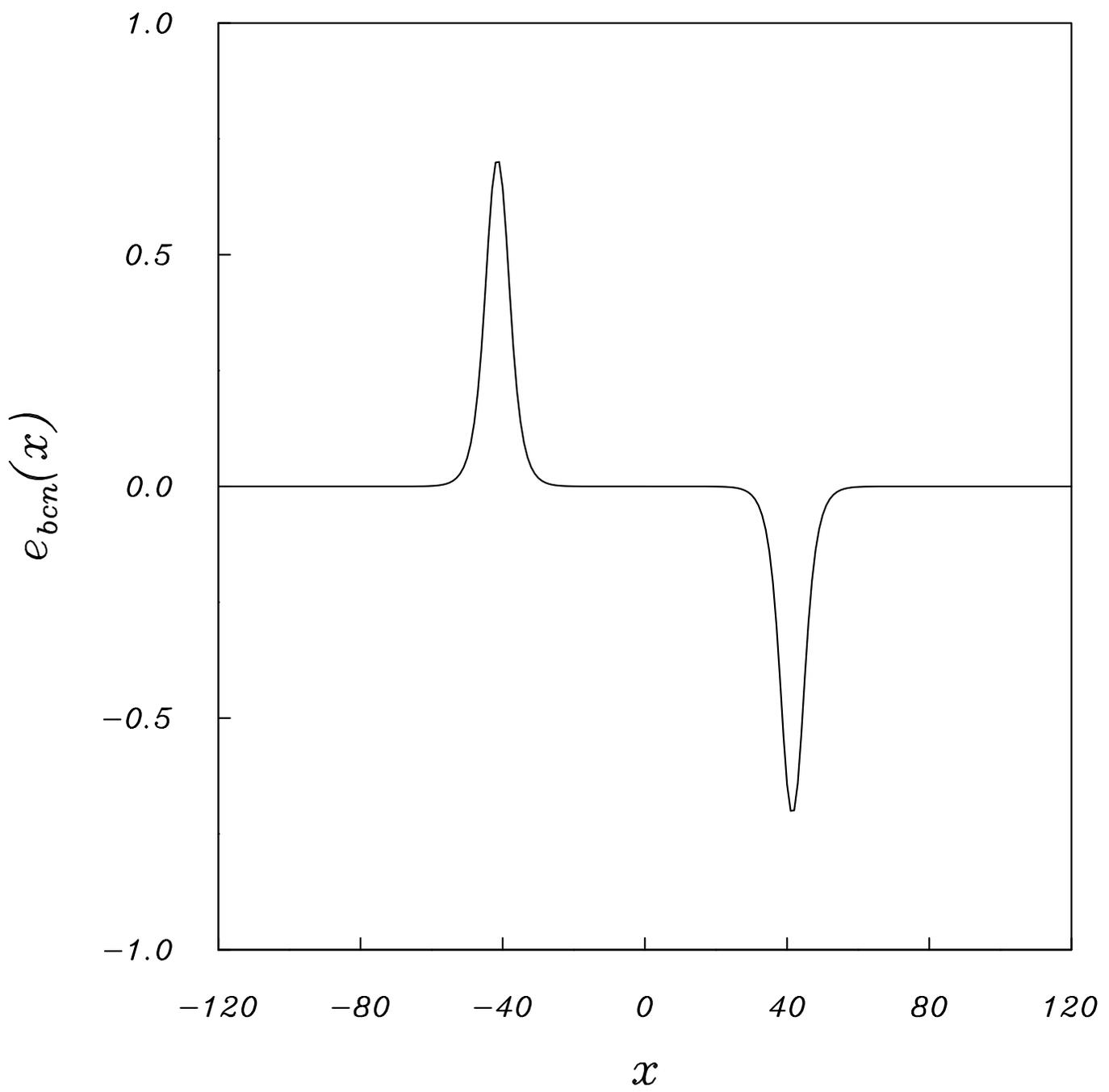
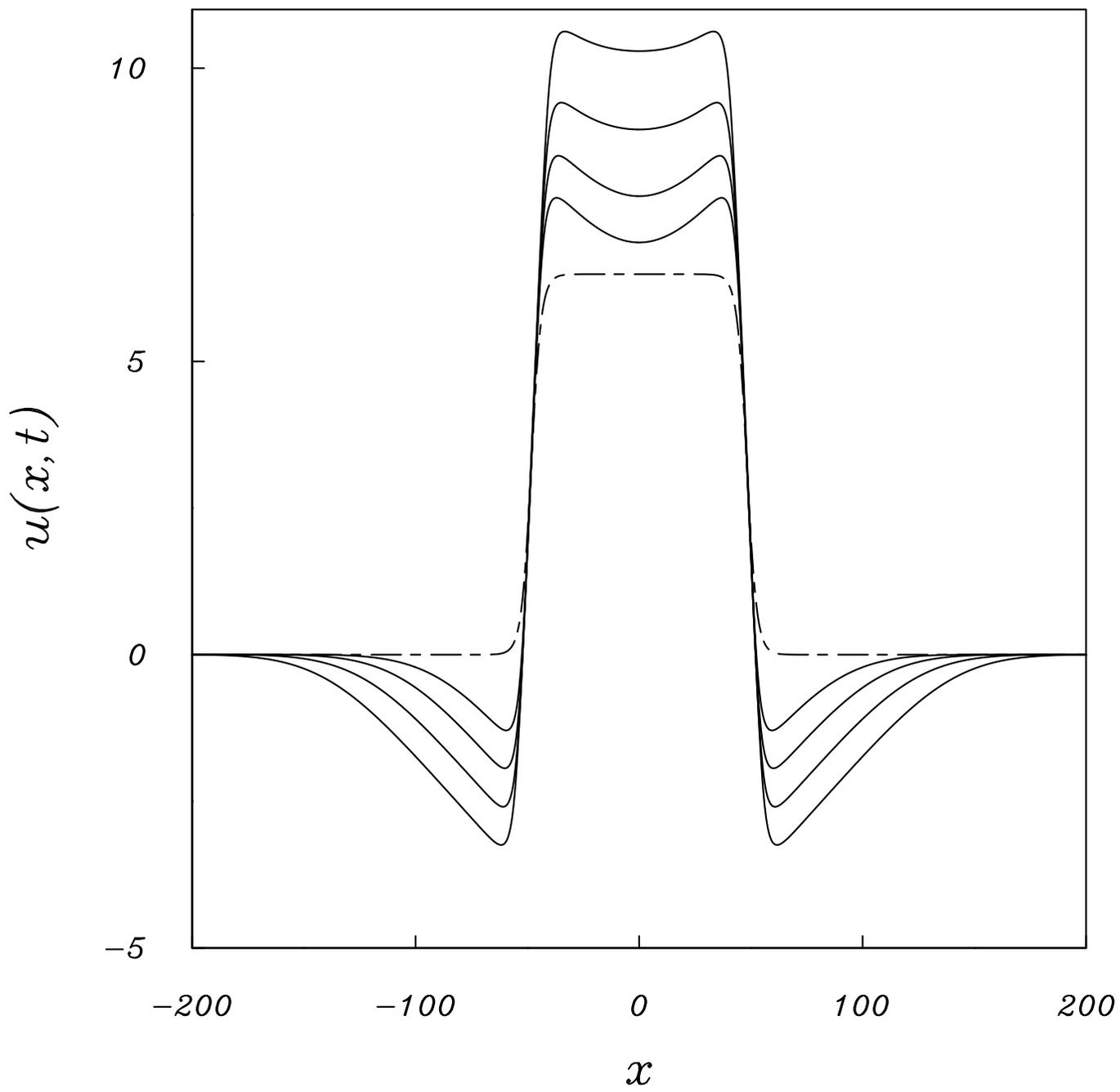
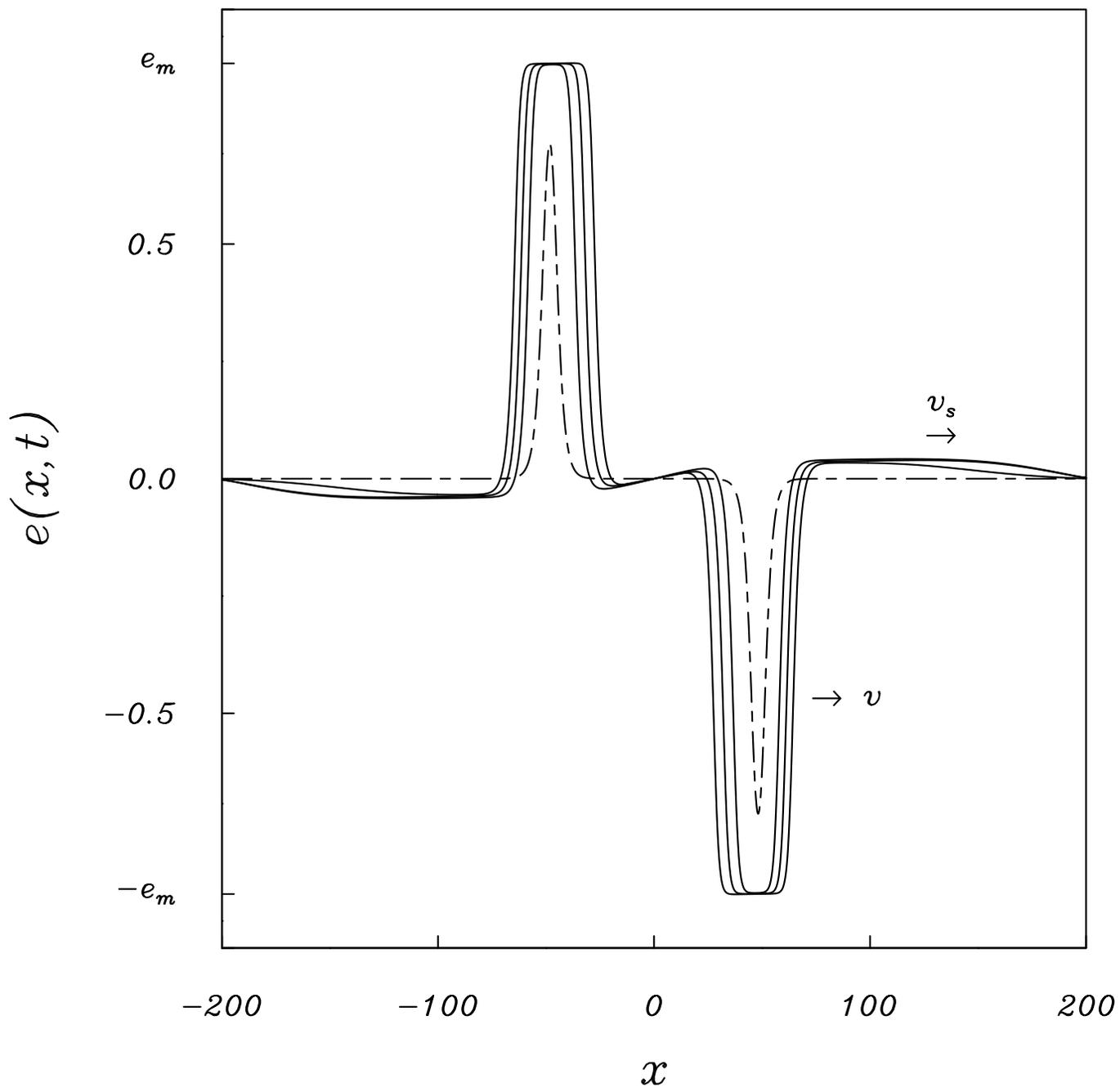
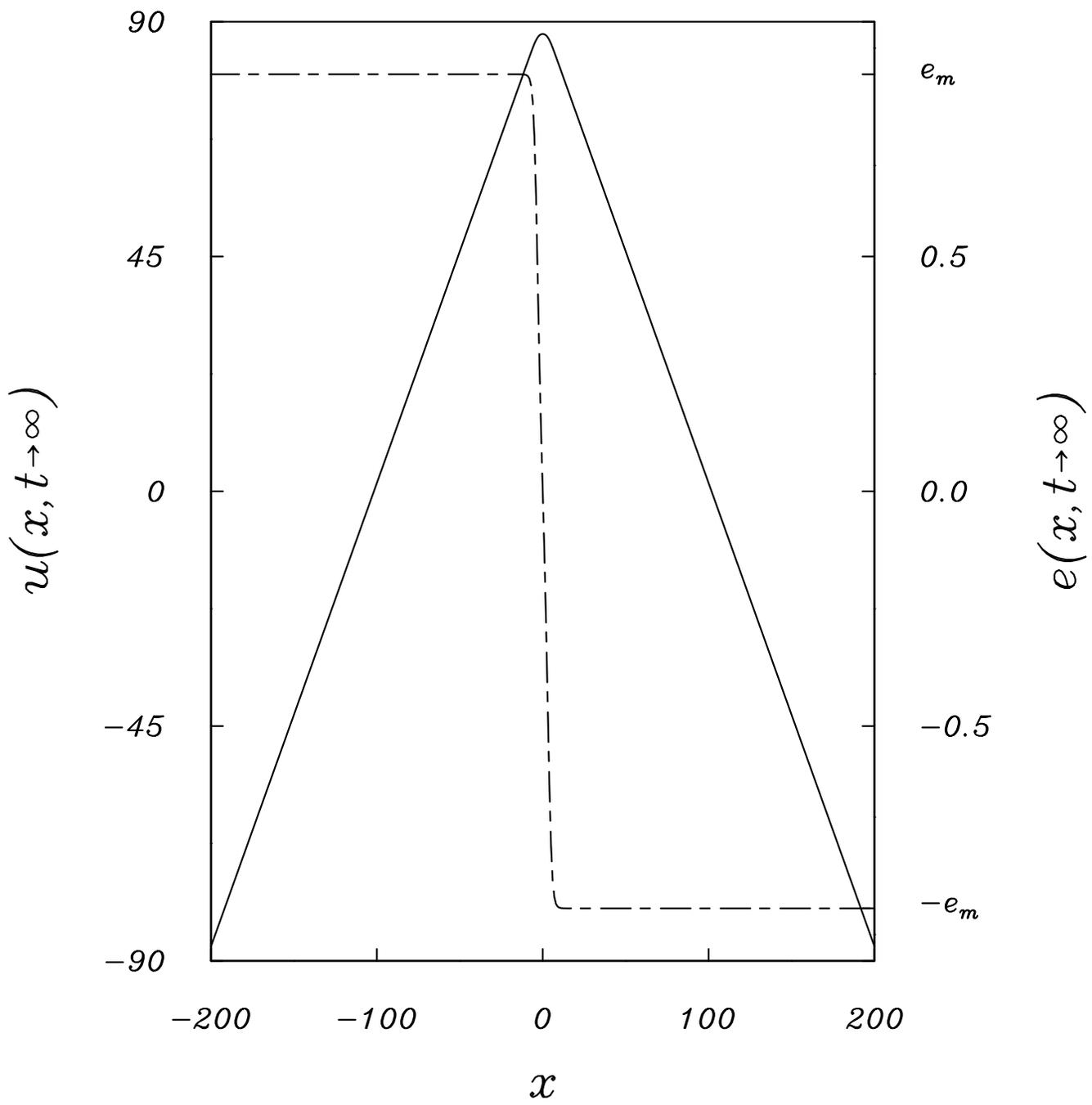
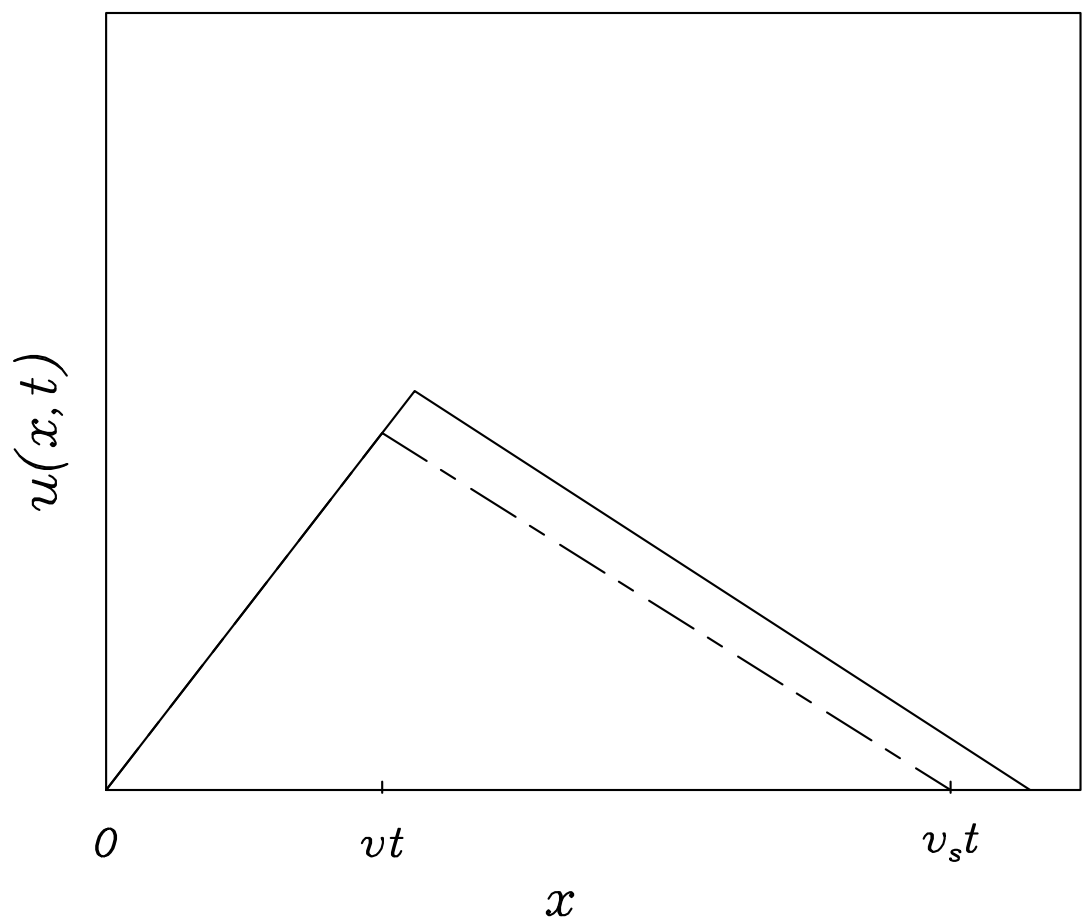
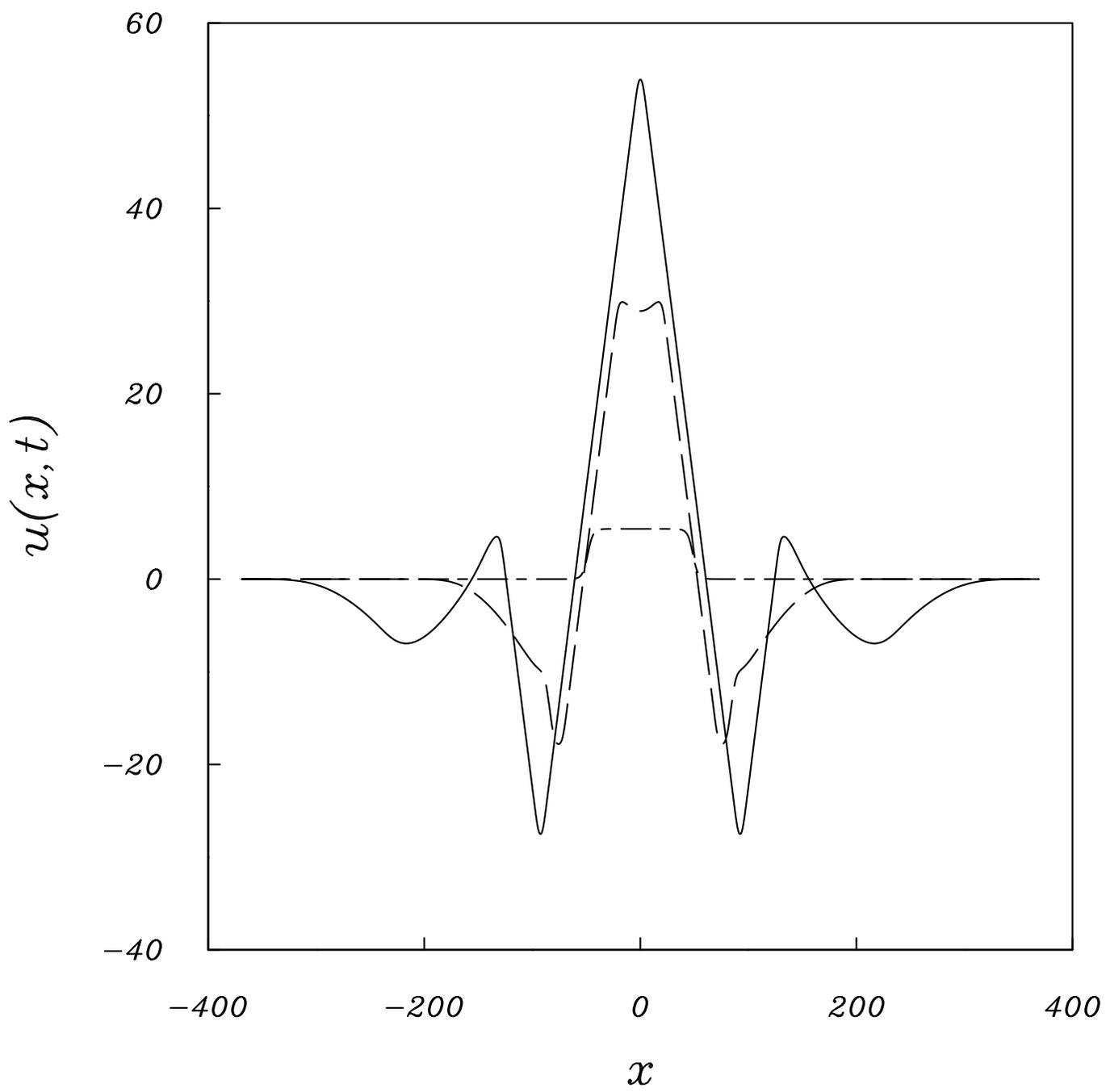
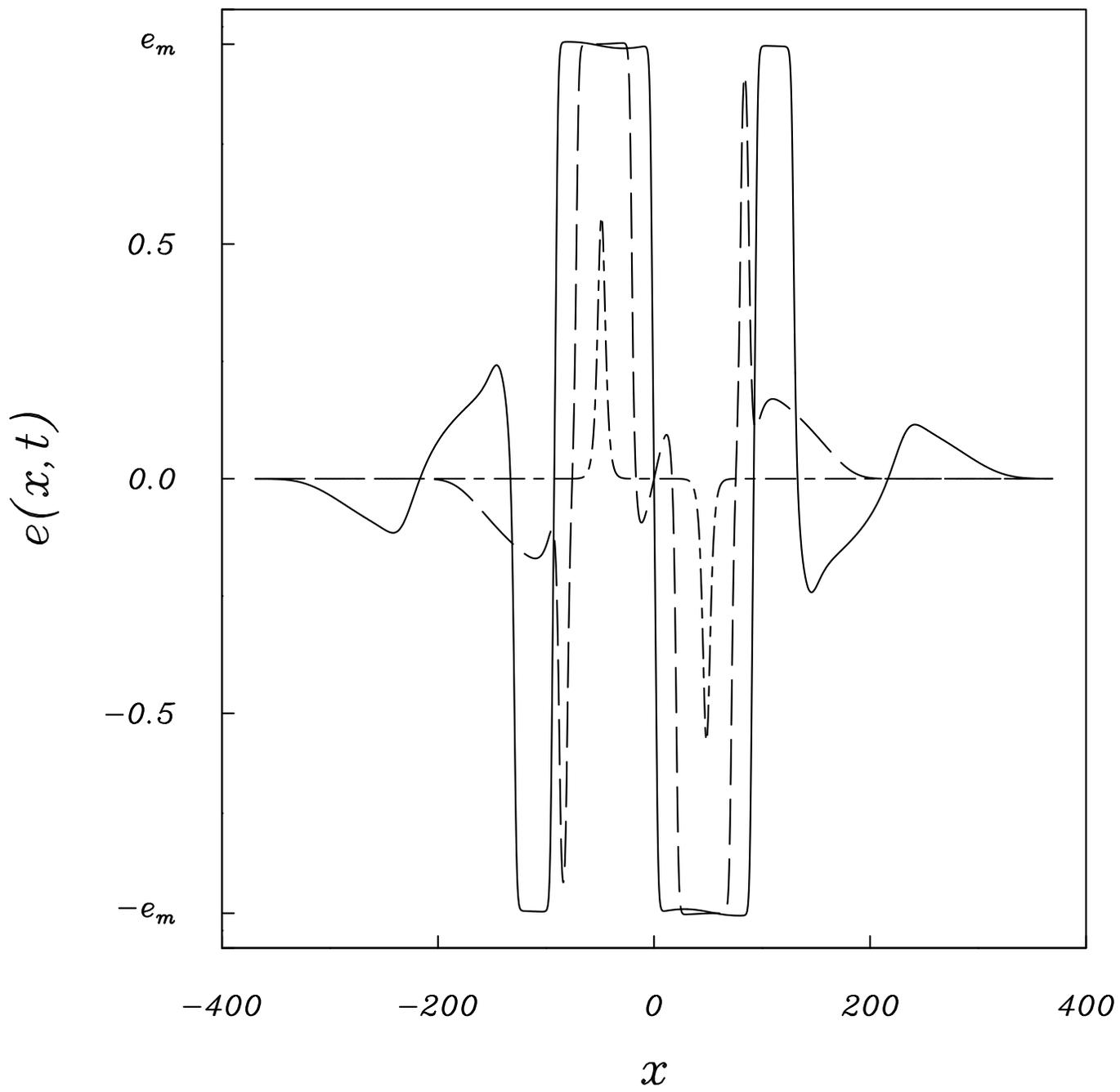
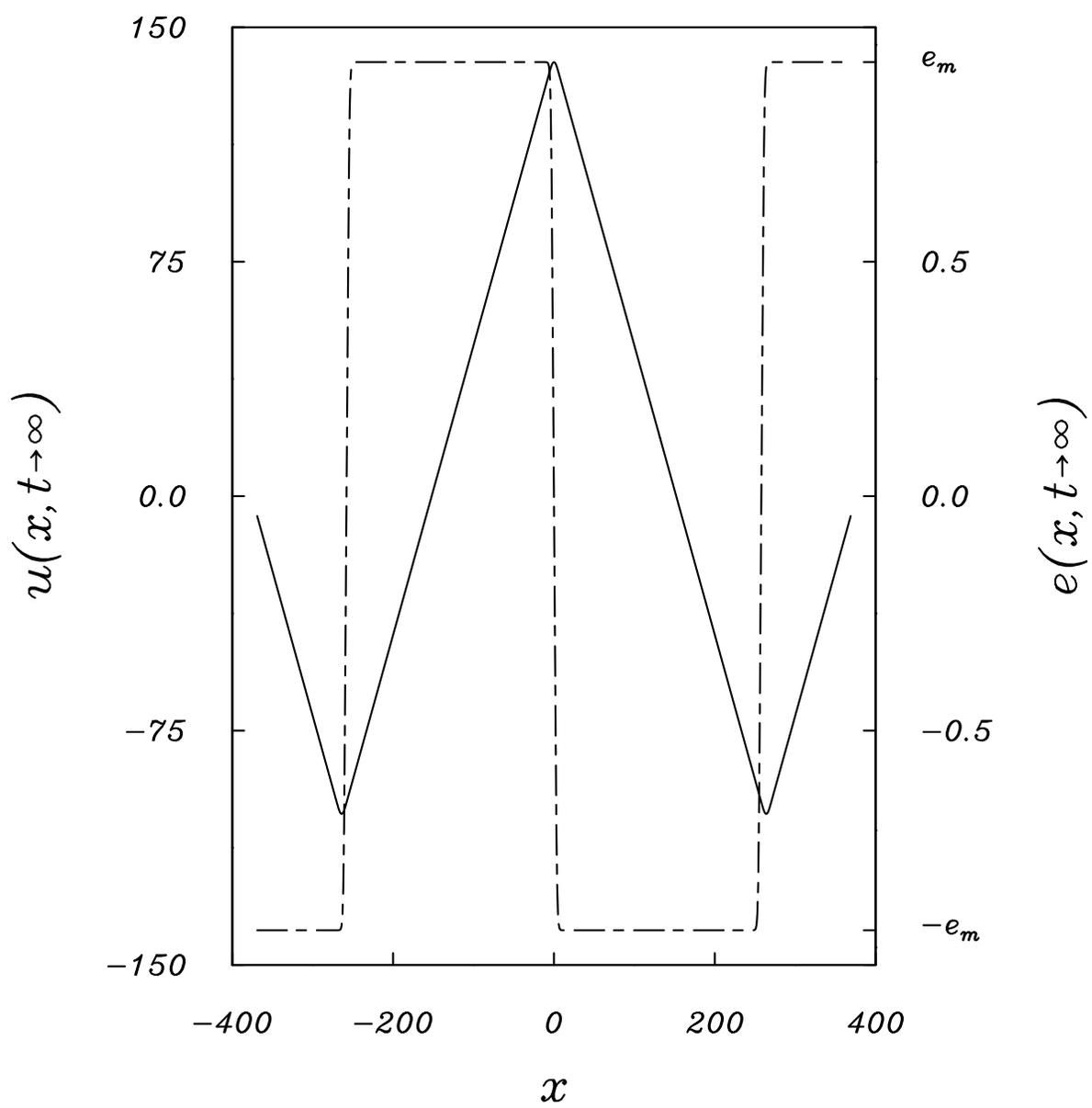
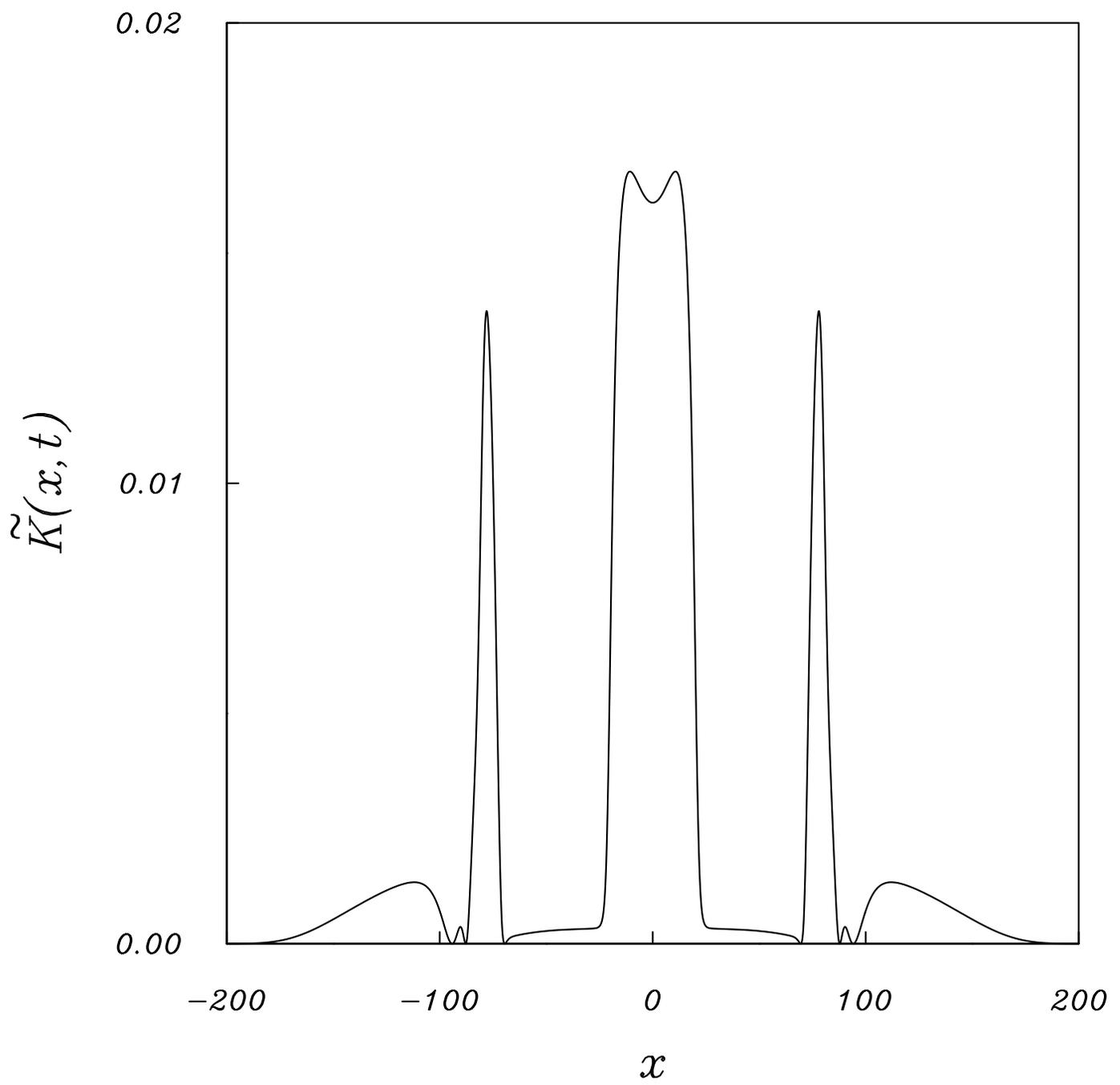
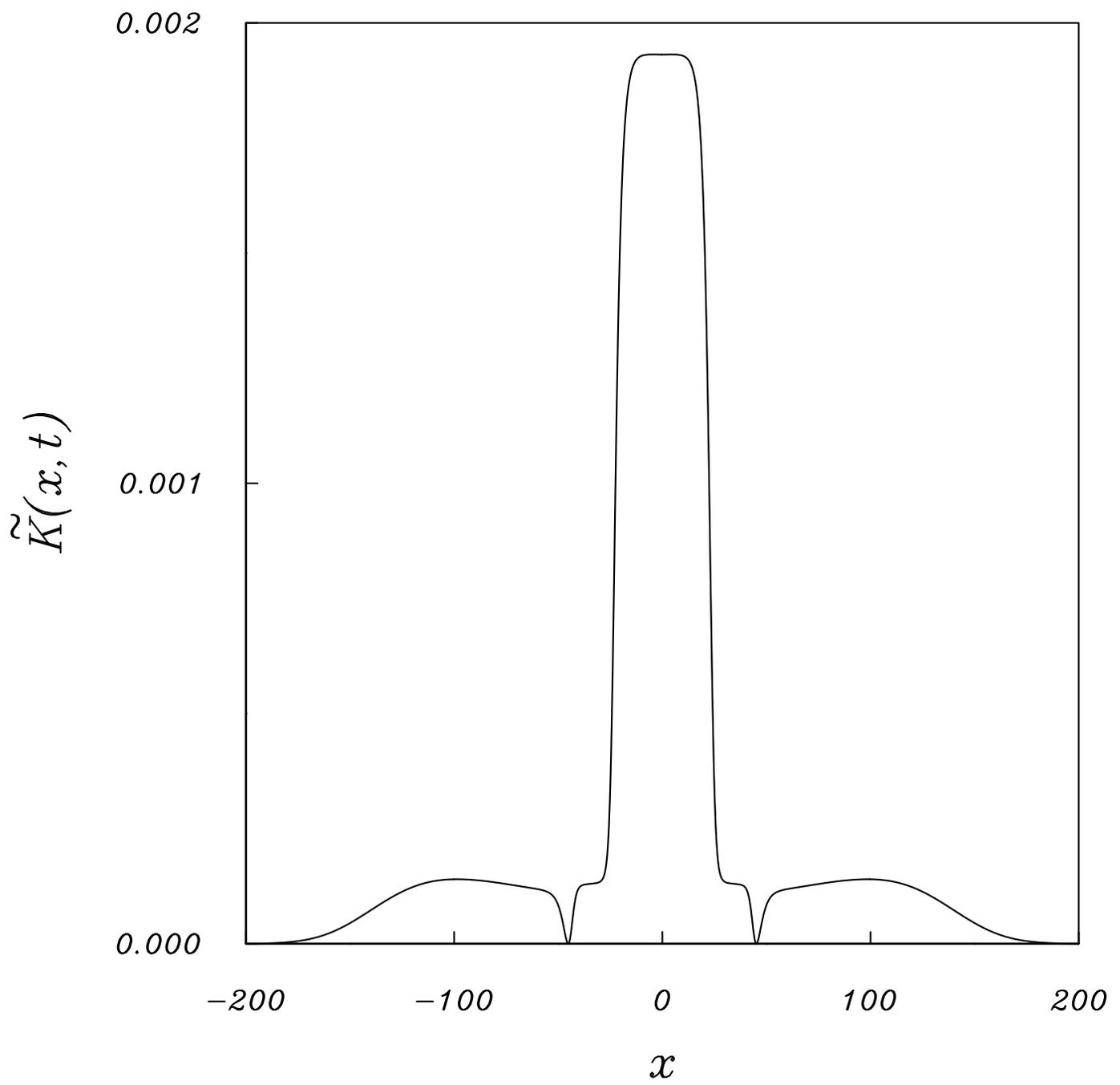
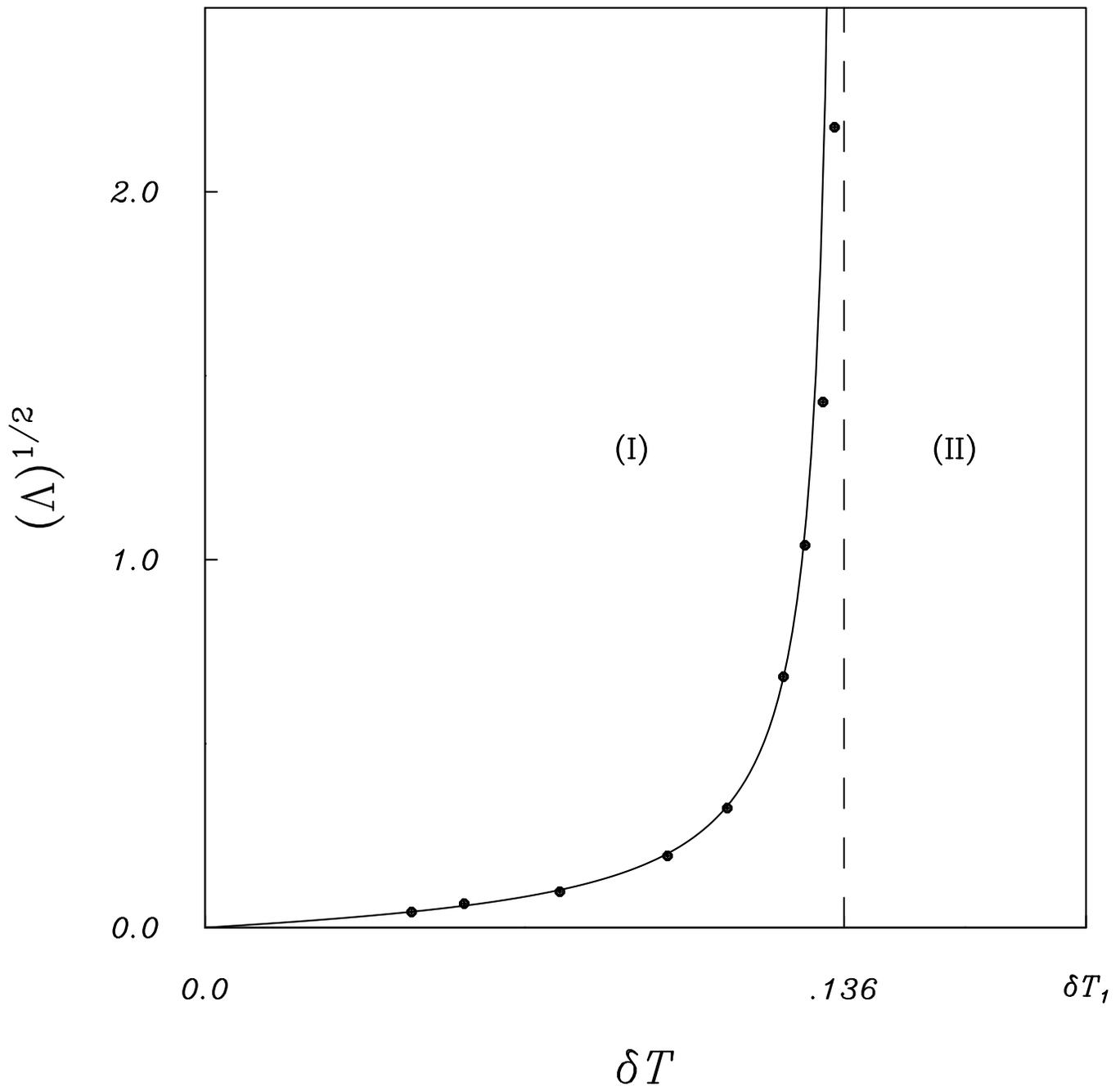
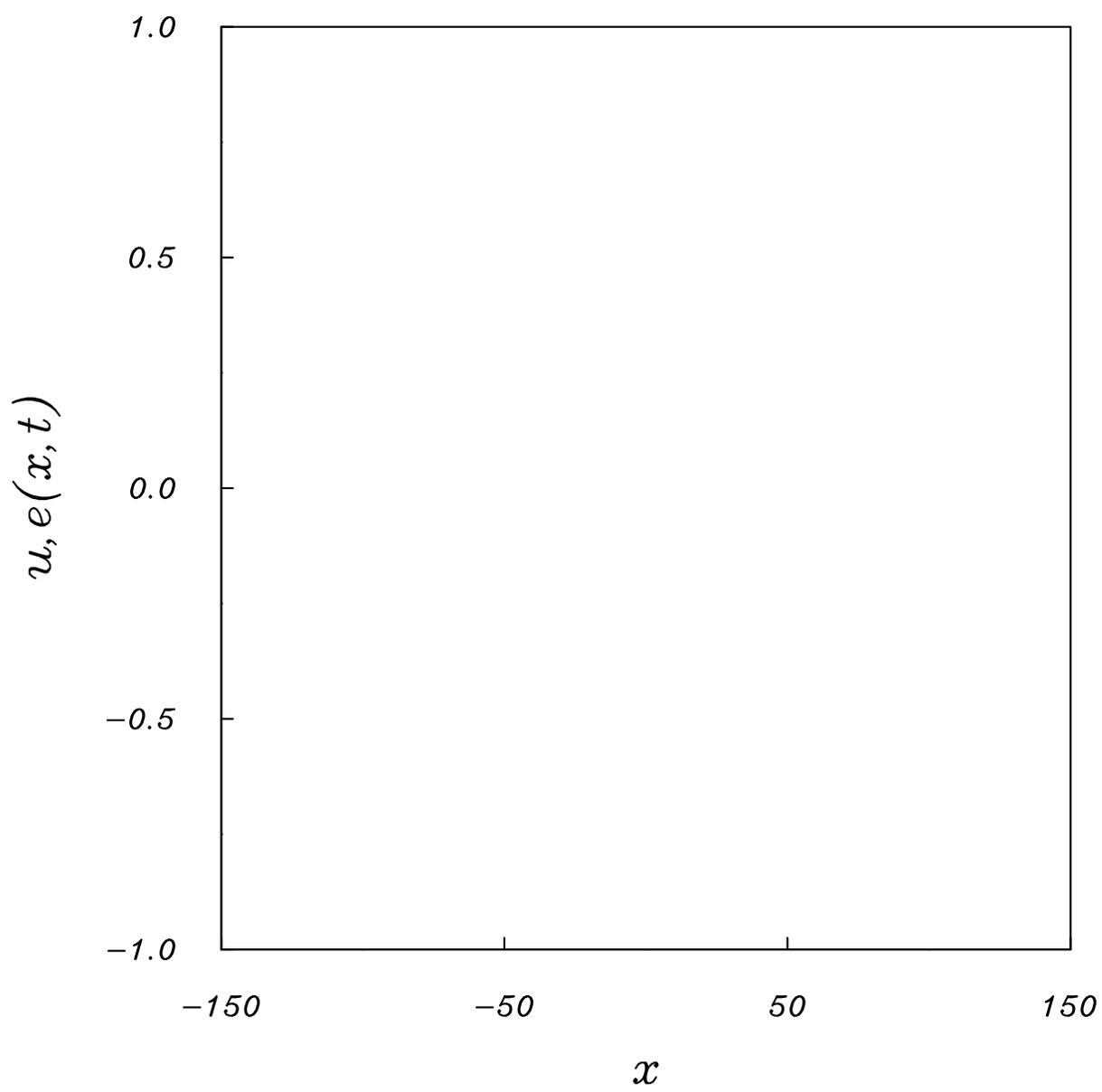
\end{document}